\documentclass[aps,prc,showpacs,showkeys,nofootinbib]{revtex4}

\usepackage{graphicx}
\usepackage{dcolumn}
\usepackage{amsthm}
\usepackage{bm}

\begin{document}

\title{New quasibound states of the compound nucleus in $\alpha$-particle capture by the nucleus
}


\author{Sergei~P.~Maydanyuk$^{(1,2)}$}%
\email{maidan@kinr.kiev.ua}%
%
\author{Peng-Ming~Zhang$^{1}$}%
\email{zhpm@impcas.ac.cn} %
\author{Li-Ping~Zou$^{1}$}%
\email{zoulp@impcas.ac.cn} %
\affiliation{$(1)$Institute of Modern Physics, Chinese Academy of Sciences, Lanzhou, 730000, China}
\affiliation{$(2)$Institute for Nuclear Research, National Academy of Sciences of Ukraine, Kiev, 03680, Ukraine}

\date{\small\today}


\begin{abstract}
We generalize the theory of nuclear decay and capture of Gamow that is based on tunneling through the barrier
and internal oscillations inside the nucleus.
In our formalism an additional factor is obtained, which describes distribution of the wave function of the $\alpha$ particle inside the nuclear region.
We discover new most stable states (called quasibound states) of the compound nucleus (CN) formed during the capture of $\alpha$ particle by the nucleus.
With a simple example, we explain why these states cannot appear in traditional calculations of the $\alpha$ capture cross sections based on monotonic penetrabilities of a barrier, but they appear in a complete description of the evolution of the CN.
Our result is obtained by a complete description of the CN evolution, which has the advantages of
(1) a clear picture of the formation of the CN and its disintegration,
(2) a detailed quantum description of the CN,
(3) tests of the calculated amplitudes based on quantum mechanics (not realized in other approaches), and
(4) high accuracy of calculations (not achieved in other approaches).
These peculiarities are shown with the capture reaction of $\alpha + ^{44}{\rm Ca}$.
We predict quasibound energy levels and determine fusion probabilities for this reaction.
The difference between our approach and theory of quasistationary states with complex energies applied for the $\alpha$ capture is also discussed.
We show
(1) that theory does not provide calculations for the cross section of $\alpha$ capture (according to modern models of the $\alpha$ capture), in contrast with our formalism, and
(2) these two approaches describe different states of the $\alpha$ capture (for the same $\alpha$-nucleus potential).
\end{abstract}


\maketitle


\section{Introduction
\label{sec.introduction}}

A traditional way in understanding of capture of $\alpha$ particles by nuclei is based on the idea of tunneling through a potential barrier~\cite{Denisov.2009.ADNDT} (see improved formalism in Ref.~\cite{Denisov.2015.PRC}).
Evaluations of the $\alpha$-particle capture rates indicate an important role of such reactions in stars~\cite{Mohr.2000.PRC,Demetriou.2002.NPA,Rauscher.2000.NPA}.
There are intensive investigations~\cite{Denisov.2005.PHRVA,Denisov.2009.PRC.v79,Denisov.2009.PRC.v80,Denisov.2009.ADNDT} providing the most accurate potential of interactions between the $\alpha$ particles and nuclei basing on
existed experimental information of $\alpha$ decay and $\alpha$ capture.
Although approaches in determination of penetrabilities of the barrier are highly developed, there is no a generally accepted method to describe a fusion in this reaction.
In heavy-ion collisions and scattering with fragments heavier than the $\alpha$ particle, an essential attention has been focused on understanding the mechanisms of the fusion
(the current status in the experimental and theoretical investigations on this topic can be seen in the recent review~\cite{Back.2014.RMP},
also in Refs.~\cite{Birkelund.1979.PRep,Vaz.1981.PRep,Birkelund.1983.ARNPS,Beckerman.1985.PRep,Steadman.1986.ARNPS,Beckerman.1988.RPP,%
Rowley.1991.PLB,Vandenbosch.1992.ARNPS,Reisdorf.1994.JPG,Dasgupta.1998.ARNPS,%
Balantekin.1998.RMP,Liang.2005.IJMPE,Canto.2006.PRep,Keeley.2007.PPNP,Hagino.2012.PTP}).
In the case of $\alpha$ capture, the model descriptions of the fusion of the $\alpha$ particle by the target nucleus inside a nuclear region are
very simplified.
The approach of sharp angular-momentum cutoff was proposed by Glas and Mosel~\cite{Glas.1975.NPA,Glas.1974.PRC}.
Eberhard et al. proposed a relation that gives information about fusion in the $\alpha$ capture of the $^{40}{\rm Ca}$ and $^{44}{\rm Ca}$ nuclei.
They compared calculated cross sections with experimental data at selected energies~\cite{Eberhard.1979.PRL}.
Recently, a more precise way to study the $\alpha$ capture problem is proposed in Ref.~\cite{Maydanyuk.2015.NPA}.
In that paper we investigated a high-precision method (called the method of multiple internal reflections, MIR) to determine fusion in the capture of $\alpha$ particles by nuclei.
With this method, we found new parametrization of the $\alpha$-nucleus potential and fusion probabilities
(see Fig.~6, Tables~2 and B.3 in Ref.~\cite{Maydanyuk.2015.NPA}).
Error in description of experimental data is decreased by 41.72 times for $\alpha + ^{40}{\rm Ca}$ and 34.06 times for $\alpha + ^{44}{\rm Ca}$ in comparison with previous results (see Fig.~5 and Table~1 in Ref.~\cite{Maydanyuk.2015.NPA}, for details).
To date, this is the most accurate and successful approach in describing experimental data for $\alpha$ capture.
Based on our fusion probability formula (see Eqs.~(21)–-(27) and Figs.~8 and 9 in Ref.~\cite{Maydanyuk.2015.NPA}),
we predicted cross sections for the $\alpha$ capture by the nucleus $^{46}{\rm Ca}$ for future experimental tests%
\footnote{This nucleus $^{46}{\rm Ca}$ is of research interest connected with discovery of new neutron magic numbers at $N=16$ and $N=26$
(see review~\cite{Penionzhkevich.2006.EPAN} of this topic;
here the standard theory gives us only seven experimentally known neutron numbers at 2, 8, 20, 28, 50, 82, 126).
In this regard, it could be interesting for experimentalists to investigate the fusion process at the capture of the $\alpha$ particle by this nucleus
based on our predictions.
%
}.

In frameworks of existing models of $\alpha$ capture, it is assumed that a complete fusion of the $\alpha$ particle and nucleus takes place after tunneling.
Cross sections of the fusion are determined by the penetrabilities.
The dependence of the penetrability on energy of the incident $\alpha$ particle is monotonic (without any minima and maxima) at each allowed orbital momentum
(see Figs.~2 and 3 in Ref.~\cite{Maydanyuk.2015.NPA} for the capture $\alpha +\, ^{44}{\rm Ca}$).
This explains the absence of peaks in the calculated cross sections of the $\alpha$ capture
and has been confirmed by existing experimental information, because
we have the cross sections for capture of the $\alpha$ particles by the nuclei $^{40}{\rm Ca}$, $^{44}{\rm Ca}$ \cite{Eberhard.1979.PRL},
$^{59}{\rm Co}$ \cite{DAuria.1968.PR}, $^{208}{\rm Pb}$~\cite{Barnett.2000.PRC}, and $^{209}{\rm Be}$~\cite{Barnett.2000.PRC}.
This is why people assume there is no any state of possible formation of the compound-nuclear system when nucleons of the $\alpha$ particle and nucleus-target form the most stable bound nuclear system at some fixed energies.

Nowadays, two approximated approaches are very popular to determine the penetrability:
(1) the Wentzel–Kramers–Brillouin (WKB) approximation and
(2) replacing the original barrier by the inverse oscillator potential, which has solutions for the wave function (see the Hill–Wheeler approach~\cite{Hill_Wheeler.1953.PR} and Wong’s formula~\cite{Wong.1973.PRL} calculating cross sections).
Such approaches have been widely used in the basis of modern coupled-channel calculations
for the study of fusion~\cite{Balantekin.1998.RMP,Back.2014.RMP}.
However, note that the region of applicability of both approaches is only near the barrier maximum.
Both approaches completely ignore the shape of the internal nuclear region and the external tail of the potential.
The WKB approximation cannot be applied for the main region of under-barrier energies,
and the oscillator potential used in the second approach is completely different from the original barrier.
In the framework of the WKB approach, a reflection from the barrier is not defined, so we cannot apply the test of quantum mechanics to check the results obtained.
As a result, an essential part of under-barrier and above-barrier energy regions for the original barrier looks like black box and cannot be correctly studied by means of the two above approaches, and consequently the proposed result for the penetrability cannot
be tested.

By such a motivation, we approximate the original potential barrier by a number of rectangular steps, for which there are exact analytical solutions for wave functions at any energy~\cite{Maydanyuk.2015.NPA,Maydanyuk.2014.JPS}.
It turns out that approximation of this approach can be reduced up to zero by increasing the number of steps, all solutions for the wave function are convergent and fully satisfy all known tests of quantum mechanics (with an accuracy up to the first 15 digits).
We study quantum processes both for deep under-barrier energies, and energies highly above the barrier maximum (that is a problem for both approaches mentioned above).
This approach has been successfully applied for different tasks of quantum physics~\cite{Maydanyuk.2011.EPJP} with the barriers of very specific shape (note that the two approximate approaches mentioned above cannot be even applied for the proper determination of the penetrabilities of these barriers).
It allows us to study the influence of the shapes of potential outside the tunneling region on the obtained penetrability.
The analysis in Refs.~\cite{Maydanyuk.2015.NPA,Maydanyuk.2014.JPS} shows that such an influence is not small,
and in some cases can change the penetrability more then 100%
\footnote{In Fig.~1 of Ref.~\cite{Maydanyuk.2015.NPA} (and in Fig.~1 of Ref.~\cite{Maydanyuk.2014.JPS}) we shown variations of the penetrability of more than four times in dependence
of the localization of the capture point (this is the internal boundary of the potential region with the barrier, for which we calculate the
penetrability) at the same incident energy of 2~MeV of the $\alpha$ particle for the capture $\alpha + ^{44}{\rm Ca}$ at $l=0$.}.
In the framework of our approach, we find that the penetrability depends on some new parameters.
They could actually be more important than the nuclear deformations.
However, these parameters are missed in the two approximate approaches above.

According to quantum mechanics, consideration of evolution of the system up to the moment of propagation of the $\alpha$ particle through the barrier is not complete (that was analyzed in Refs.~\cite{Maydanyuk.2015.NPA,Maydanyuk.2014.JPS}).
Conservation of a flux of wave function requires us to take a further evolution of this system into account.
Our research in this paper starts from an analysis of this evolution of the compound system in that stage.
It turns out that such a consideration leads to the appearance of oscillations of this system and its disintegration (and allows us to include mechanisms of fusion).
In frameworks of unified formalism we join the tunneling processes and oscillations inside the internal nuclear region the first time.
Note that the idea introduced by Gamow in 1928 applied to describe $\alpha$ decay~\cite{Gamow.1928.ZP}, where these two processes were considered separately (and there is no approach combining these two processes) for determination of half-lives of decay.
Until now, half-lives of the $\alpha$ decay of nuclei are determined on such a basis with inclusion of spectroscopic factors
(see, for example, Refs.~\cite{Buck.1993.ADNDT,Akovali.1998.NDS,Duarte.2002.ADNDT,
Audi.2003.NPA,Dasgupta-Schubert.2007.ADNDT,Silisteanu.2012.ADNDT,Lovas.1998.PRep,Sobiczewski.2007.PPNP,
Stewart.1996.NPA,Xu.2006.PRC,Nazarewicz.2012.PRC,Delion.2013.PRC,Silisteanu.2015.RJP,www_library}).

Another implication of our method shown in this paper is the appearance of the maximally stable states of the compound system at some energies of the incident $\alpha$ particle (at monotonic penetrabilities of the barrier).
The existence of such maximally stable states (we call them as quasibound) reflects the quantum nature of collisions of nuclei; however, it cannot
be explained by traditional methods (see, for example, methods based on Ref.~\cite{Eberhard.1979.PRL} for comparison).
In this regard, new questions will appear.
By how much do oscillations prevail, how fast does the fusion takes place, and in which space region does the fusion dominate?
To clarify these questions, in this paper we improve the method proposed in Ref.~\cite{Maydanyuk.2015.NPA}
by including a new formalism of evolution of the compound system (after tunneling) with possible fusion.
\section{Method
\label{sec.method}}

To clearly understand how the quasibound states of the compound-nuclear system appear with monotonic penetrabilities, let us consider the simplest picture of scattering of an $\alpha$ particle off a nucleus in a spherically symmetric scenario.
It turns out that the simplest potential applicable for this aim and its corresponding general solution
of the wave function (up to normalization) are
%
%
\begin{equation}
\begin{array}{lll}
  V(r) = \left\{
  \begin{array}{cll}
    V_{1}   & \mbox{at } r_{\rm min} < r \leq r_{1}  & \mbox{(region 1)}, \\
    0   & \mbox{at } r_{1} \leq r \leq r_{\rm max}   & \mbox{(region 2)},
  \end{array} \right. &
\psi(r, \theta, \varphi) = \displaystyle\frac{\chi(r)}{r} Y_{lm}(\theta, \varphi), &
\chi(r) = \left\{
\begin{array}{lll}
   \alpha_{1}\, e^{ik_{1}r} + \beta_{1}\, e^{-ik_{1}r}  & \mbox{(region 1)}, \\ 
   e^{-ik_{2}r} + A_{R}\,e^{ik_{2}r} & \mbox{(region 2)}, 
\end{array} \right.
\end{array}
\label{eq.2.1.1}
\end{equation}
where $V_{1}<0$,
$r_{\rm min} \ge 0$,
$\alpha_{1}$, $\beta_{1}$ and $A_{R}$ are unknown amplitudes, $Y_{lm}(\theta ,\varphi)$ is the spherical function, and $k_{j} = \frac{1}{\hbar}\sqrt{2m(E-V_{j})}$ are complex wave numbers ($j=1,2$, $V_{2}=0$).
We fix the normalization of the wave function so that the modulus of the amplitude of the incident wave $e^{-ik_{2}r}$ equals unity.

According to the Multiple Internal Reflection (MIR) method in Ref.~\cite{Maydanyuk.2015.NPA} (see Refs.~\cite{Maydanyuk.MIR.all} also, for details), the scattering of the particle on the potential is sequentially considered by steps of propagation of a wave packet relative to each boundary.
In the first step we consider a wave $e^{-ik_{2}r}$ in region 2, which is incident on the boundary at point $r_{1}$.
This wave is transformed into a new wave $\beta_{1}^{(1)} e^{-ik_{1}r}$ propagated to the center in region 1, and
a new wave $\alpha_{2}^{(1)} e^{ik_{2}r}$ reflected from the boundary and propagated into region 2.
We have such a wave function for this process:
\begin{equation}
\chi^{(1)}(r) = \left\{
\begin{array}{lll}
   \beta_{1}^{(1)}\, e^{-ik_{1}r}  & \mbox{at } r_{\rm min} < r \leq r_{1}, \\
   e^{-ik_{2}r} + \alpha_{2}^{(1)}\,e^{ik_{2}r} & \mbox{at } r_{1} \leq r \leq r_{\rm max}.
\end{array} \right.
\label{eq.2.1.1.1}
\end{equation}
The transmitted wave is formed in the internal nuclear region.
Thus, it describes the formation of a compound nucleus and its further evolution.
The reflected wave describes reflection of the particle by Coulomb forces of the nucleus.
Therefore, in the framework of this extremely simple scheme,
we have separated the scattering of particle off the nucleus into two physically different processes:
(1) formation of the compound nucleus and its possible disintegration and
(2) the potential scattering without compound-nucleus formation.

In the second step we consider the wave $\beta_{1}^{(1)}\, e^{-ik_{1}r}$ in the region 1 formed in the previous step.
This wave propagates to center of the nucleus and is transformed into a new wave $\alpha_{1}^{(2)}\,e^{ik_{1}r}$.
In the third step, we consider the wave $\alpha_{1}^{(2)}\,e^{ik_{1}r}$ which is incident on the boundary at $r_{1}$ and transformed into
a new wave in region 2 (describing transmission through the boundary) which propagates outside,
and another new wave in region 1 (describing reflection from the boundary) which propagates to center.
One can describe these processes by wave functions:
\begin{equation}
\begin{array}{llll}
  \chi^{(2)}(r) =
  \beta_{1}^{(1)}\, e^{-ik_{1}r} + \alpha_{1}^{(2)}\,e^{ik_{1}r} & \mbox{at } r_{\rm min} < r \leq r_{1}; &
\chi^{(3)}(r) = \left\{
\begin{array}{lll}
   \alpha_{1}^{(2)}\,e^{ik_{1}r} + \beta_{1}^{(3)}\, e^{-ik_{1}r}  & \mbox{at } r_{\rm min} < r \leq r_{1}, \\
   \alpha_{2}^{(3)}\,e^{ik_{2}r} & \mbox{at } r_{1} \leq r \leq r_{\rm max}.
\end{array} \right.
\end{array}
\label{eq.2.1.1.4}
\end{equation}
Here, $\alpha_{j}^{(i)}$ and $\beta_{j}^{(i)}$ are unknown amplitudes (we add upper index $i$ denoting number of step, and bottom index $j$ denoting number of region).
We find the following recurrent relations from conditions of continuity of the full wave function and its derivative:
\begin{equation}
\begin{array}{lllll}
  \alpha_{2}^{(1)} = R_{1}^{-}, &
  \beta_{1}^{(1)} = T_{1}^{-}, &
  \alpha_{1}^{(2)} = R_{0}\, \beta_{1}^{(1)}, &
  \alpha_{2}^{(3)} = \alpha_{1}^{(2)}\, T_{1}^{+}, &
  \beta_{1}^{(3)} = \alpha_{1}^{(2)}\, R_{1}^{+}, \\
%
%
\vspace{1mm}
  R_{1}^{-}\hspace{-0.8mm} = \hspace{-1mm}\displaystyle\frac{k-k_{1}}{k+k_{1}}\, e^{-2ikr_{1}}, &
  T_{1}^{-}\hspace{-0.8mm} = \hspace{-1mm}\displaystyle\frac{2k}{k+k_{1}}\, e^{-i(k-k_{1})r_{1}}, &
  R_{0}\hspace{-0.6mm} = - e^{-2ik_{1}r_{\rm min}}, &
  T_{1}^{+}\hspace{-0.8mm} = \hspace{-1mm}\displaystyle\frac{2k_{1}}{k+k_{1}}\, e^{i(k_{1}-k)r_{1}}, &
  R_{1}^{+}\hspace{-0.8mm} = \hspace{-1mm}\displaystyle\frac{k_{1}-k}{k+k_{1}}\, e^{2ik_{1}r_{1}}.
\end{array}
\label{eq.2.1.1.6}
\end{equation}
Here, 
we introduce new amplitudes $T_{1}^{-}$ and $R_{1}^{+}$, describing transmission and reflection concerning the boundary
(bottom index ``1'' or ``0'' indicates number of the boundary, upper sign ``$-$'' or ``$+$'' indicates
the negative or positive radial direction, respectively, of the incident wave in determination of the amplitude).
Each next step in such a consideration for propagation of waves is similar to one of these three steps.
With the above analysis we find recurrent relations for new unknown amplitudes
and calculate the following summations of all waves:
\begin{equation}
\begin{array}{llll}
  \displaystyle\sum\limits_{i=1} \beta_{1}^{(i)} =
  A_{\rm osc}\, T_{1}^{-}, &

  \displaystyle\sum\limits_{i=1} \alpha_{1}^{(i)} =
  R_{0}\, \displaystyle\sum\limits_{i=1} \beta_{1}^{(i)}, &


  \displaystyle\sum\limits_{i=2} \alpha_{2}^{(i)} =
  A_{\rm osc}\, T_{1}^{-} R_{0} T_{1}^{+}, &

  A_{\rm osc} =
  \Bigl( 1 + \displaystyle\sum\limits_{i=1} (R_{0} R_{1}^{+})^{i} \Bigr) =
  \displaystyle\frac{1}{1 - R_{0} R_{1}^{+}}.
\end{array}
\label{eq.2.1.3.1}
\end{equation}
%
%
%
The factor $A_{\rm osc}$ describes oscillation of waves inside internal region 1
(so we call it the \emph{amplitude of oscillations}).
At $R_{0}=-1$ we obtain
\begin{equation}
\begin{array}{lll}
  \displaystyle\sum\limits_{i=1} \beta_{1}^{(i)} \hspace{-0.5mm} =
  -\hspace{-1mm} \displaystyle\sum\limits_{i=1} \alpha_{1}^{(i)} =
  A_{\rm osc} \displaystyle\frac{2k e^{-i(k-k_{1})r_{1}}} {k+k_{1}}, &

  \displaystyle\sum\limits_{i=2} \alpha_{2}^{(i)} =
    - A_{\rm osc} \displaystyle\frac{4kk_{1} e^{2i(k_{1}-k)r_{1}}}{(k+k_{1})^{2}}, &

  A_{\rm osc} = \displaystyle\frac{k + k_{1}}{(k+k_{1}) + (k_{1}-k)\,e^{i2k_{1}r_{1}}}.
\end{array}
\label{eq.2.1.3.11}
\end{equation}
Note that full flux of all outgoing waves equals the flux of incident wave ($k$ and $k_{1}$ are real):
$\bigl|\, \alpha_{2}^{(1)} + \sum\limits_{i=2} \alpha_{2}^{(i)} \bigr|^{2} = 1$.

\begin{figure}[htbp]
\centerline{\includegraphics[width=86mm]{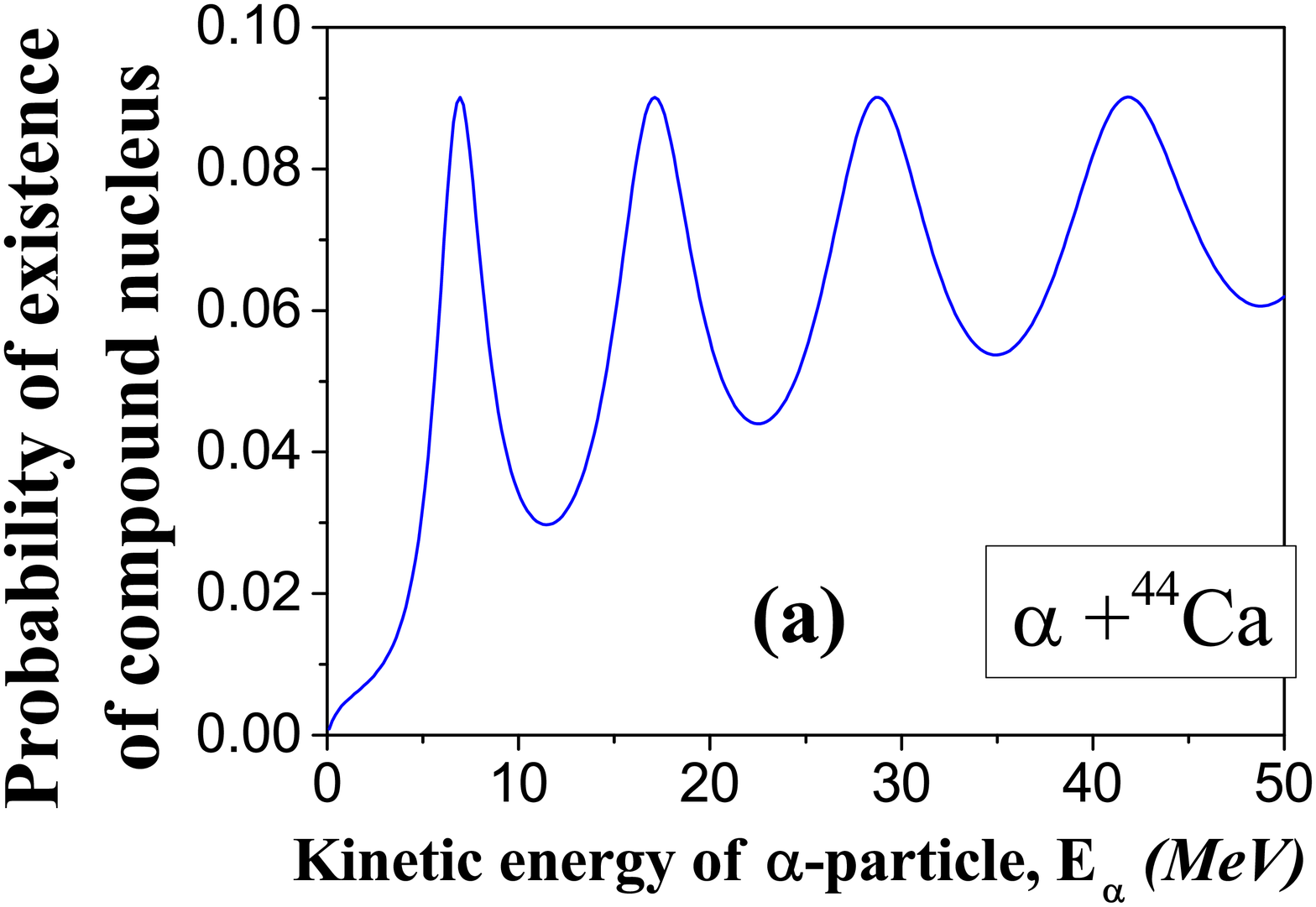}
\hspace{-5mm}\includegraphics[width=86mm]{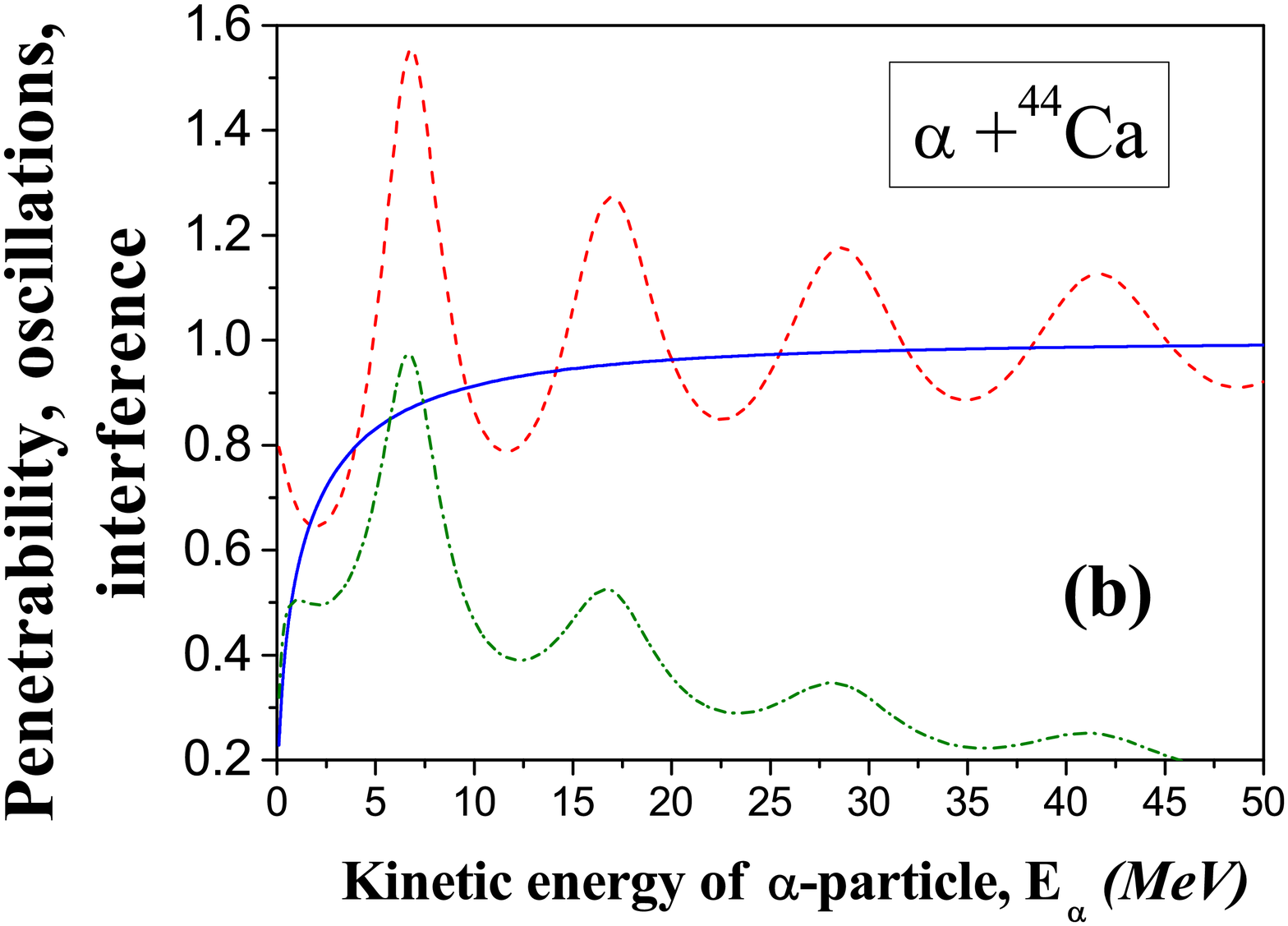}
}
\vspace{-3mm}
\caption{\small (Color online)
Probability of existence of the compound nucleus $P_{\rm cn}$ (a) defined by Eq.~(\ref{eq.2.1.5.4}),
penetrability of the boundary $T_{\rm bar}$ (b, blue solid line) defined by Eq.~(\ref{eq.2.1.5.4}),
modulus of the amplitude of oscillations $A_{\rm osc}$ (b, red dashed line) defined in Eq.~(\ref{eq.2.1.3.11}),
interference term $P_{\rm interf}$ (b, green dash-dotted line) defined in Eq.~(\ref{eq.2.1.6.2}),
in dependence on the energy of the incident $\alpha$ particle
for reaction $\alpha + ^{44}{\rm Ca}$ at $l=0$
(parameters $V_{1}$ and $r_{1}$ are taken concerning to depth of well and coordinate of maximum of the realistic radial barrier at parametrization~\cite{Denisov.2005.PHRVA} for this reaction:
we obtain $V_{1}=-23.73$~MeV and $r_{1}=8.935$~fm;
in all presented calculations test is fulfilled with coincidence of the first 14 digits).
One can see a clear presence of maximums of the probability of existence of the compound nucleus,
the amplitude of oscillations and interference term
(energies of maximums of these functions are very close, but not coincident;
these functions are principally different near zero energy),
whereas the penetrability is a smooth monotonous function.
Without inclusion of function describing internal processes,
the cross-section of fusion defined only on the basis of the penetrabilities of the barrier
(for example, as in approach~\cite{Eberhard.1979.PRL})
cannot give information about these maximally stable states of compound nucleus.
Factor $P_{\rm cn}$ has the same maximums,
its oscillatory behavior is explained mainly by the amplitude of oscillations $A_{\rm osc}$.
\label{fig.2.1.1}}
\end{figure}

Let us calculate integral of the square of the modulus of the wave function over the region 1 (at $R_{0}=-1$):
\begin{equation}
\begin{array}{llllll}
  P_{\rm cn} = \displaystyle\int\limits_{0}^{r_{1}} |\varphi(r)|^{2}\; dr =
    P_{\rm osc}\, T_{\rm bar}\, P_{\rm loc}, &
  P_{\rm osc} = |A_{\rm osc}|^{2}, &
  T_{\rm bar} \equiv \displaystyle\frac{k_{1}}{k_{2}}\; \bigl| T_{1}^{-} \bigr|^{2}, &
  P_{\rm loc} = 2\, \displaystyle\frac{k_{2}}{k_{1}}\; \Bigl( r_{1} - \displaystyle\frac{\sin(2k_{1}r_{1})}{2k_{1}} \Bigr).
\end{array}
\label{eq.2.1.5.4}
\end{equation}
This integral is interpreted as the probability of existence of the compound nucleus formed (in space region up to $r_{1}$) during the scattering.
One can see from Fig.~\ref{fig.2.1.1} that this probability depends on the energy of $\alpha$ particle and it has maxima and minima (for the same fixed normalization of the incident wave).
This is because $P_{\rm cn}$ is the explicit multiplication of the penetrability $T_{\rm bar}$, coefficient of oscillations $P_{\rm osc}$, and one additional new factor $P_{\rm loc}$.
So, we have obtained a generalization of Gamow's idea in determination of half-life of nuclear decay, that based on the penetrability of barrier and internal oscillations inside the internal region.
But here we obtain also a new factor $P_{\rm loc}$, which can be interpreted as space distribution of the $\alpha$ particle
inside the nuclear region (at one oscillation) described via the wave function.
We call it the ``coefficient of localization''.


Moreover, there is an interference term between the wave reflected in the first step from the boundary $r_{1}$ (describing the potential scattering without compound-nucleus formation) and summation of all other waves outgoing to region 2
(which are formed in formation of the compound nucleus and its decay).
We have ($R_{0} = -1$):
\begin{equation}
\begin{array}{lll}
  P_{\rm interf} \equiv
  2\; | \alpha_{2}^{(1)*} \cdot \displaystyle\sum\limits_{i=2} \alpha_{2}^{(i)})| & = &
  \displaystyle\frac{4\sqrt{2}\,kk_{1}\, |k-k_{1}|}{(k+k_{1})^{2}}\,
  \displaystyle\frac{1}{\sqrt{ k^{2} (1 -\cos (2k_{1}r_{1})) + k_{1}^{2}\,(1 + \cos (2k_{1}r_{1})} }.
\end{array}
\label{eq.2.1.6.2}
\end{equation}
For instance, in Fig.~\ref{fig.2.1.1} we present the coefficients for the reactions of $\alpha + ^{44}{\rm Ca}$ at $l=0$.
%
%
%


The complete fusion could be described via a requirement:
\emph{the flux of each wave propagating inside the nuclear region is suppressed up to zero}.
Mathematically, this condition can be realized by
\begin{equation}
\begin{array}{lllllll}
  R_{0} \to 0, &
  \displaystyle\sum\limits_{i=1} \beta_{1}^{(i)} = T_{1}^{-}, &
  \displaystyle\sum\limits_{i=1} \alpha_{1}^{(i)} = 0, &
  \displaystyle\sum\limits_{i=2} \alpha_{2}^{(i)} = 0, &
  A_{\rm osc} = 1, &
  P_{\rm cn} = \displaystyle\frac{k_{2}\, r_{1}}{k_{1}}\, T_{\rm bar}.
\end{array}
\label{eq.2.1.7.2}
\end{equation}
According to the obtained amplitudes, this fusion is fast and complete.
It takes place from the moment after waves tunneling the barrier, and there are no further oscillations of waves inside nuclear region.
If we construct the compound nucleus without fast complete fusion,
we should partially suppress fluxes inside the internal region, i.e. it needs to make condition (\ref{eq.2.1.7.2}) less strict.
Thus, we introduce a new coefficient $p_{1}$ in region 1 and redefine amplitude $R_{0}$ as
\begin{equation}
\begin{array}{lll}
  R_{0}^{\rm (new)} \equiv R_{0}^{\rm (old)} \cdot (1 - p_{1}), &
  0 \le p_{1} \le 1.
\end{array}
\label{eq.2.1.7.7}
\end{equation}
At $p_{1}=1$ Eq.~(\ref{eq.2.1.7.7}) is transformed to Eq.~(\ref{eq.2.1.7.2}) and fast complete fusion is obtained, while at $p_{1}=0$ we have the compound nucleus without fusion.



Now we would like to generalize the idea presented above for a realistic potential of $\alpha$ capture with barrier of complicated shape, which has successfully been approximated by a sufficiently large number $N$ of rectangular steps
in Ref.~\cite{Maydanyuk.2015.NPA} (see logic of this method, tests, details, and reference therein).
In addition to our previous study~\cite{Maydanyuk.2015.NPA},
in this paper we consider that the capture of the $\alpha$ particle by the nucleus takes place in a region with number $N_{\rm cap}$ after its tunneling through the barrier from the right part of potential
and next propagations of all waves along the potential
are possible, which follows from conservation of full flux based on full wave function.
A general solution of the radial wave function (up to its normalization) for the above-barrier energies has the form (6)--(7) from Ref.~\cite{Maydanyuk.2015.NPA}.
We have fixed a normalization of the wave function so that the modulus of the amplitude of the incident wave $e^{-ik_{N}r}$ (in region with number $N$) equals unity.
We shall search a solution of the unknown amplitudes of the wave function by the MIR approach.
Each step in such a consideration of packet propagation is similar to one of the first independent $2N-1$ steps.
Amplitudes $T_{1}^{\pm}$ \ldots $T_{N-1}^{\pm}$ and $R_{1}^{\pm}$ \ldots $R_{N-1}^{\pm}$ are expressed as
\begin{equation}
\begin{array}{llll}
   T_{j}^{+} = \displaystyle\frac{2k_{j}}{k_{j}+k_{j+1}} \,e^{i(k_{j}-k_{j+1}) r_{j}}, &
   T_{j}^{-} = \displaystyle\frac{2k_{j+1}}{k_{j}+k_{j+1}} \,e^{i(k_{j}-k_{j+1}) r_{j}}, &
   R_{j}^{+} = \displaystyle\frac{k_{j}-k_{j+1}}{k_{j}+k_{j+1}} \,e^{2ik_{j}r_{j}}, &
   R_{j}^{-} = \displaystyle\frac{k_{j+1}-k_{j}}{k_{j}+k_{j+1}} \,e^{-2ik_{j+1}r_{j}}.
\end{array}
\label{eq.2.2.1.3}
\end{equation}

Now, let us find a wave propagating to the left in region with number $j-1$, which is formed after transmission through the boundary $r_{j-1}$ of all possible incident waves,
produced  as result of all possible reflections and transmissions of any waves in the right part of potential from this boundary.
The amplitude of this wave can be determined as a summation of the amplitudes of all waves incident on boundary at point $r_{j-1}$ multiplied by a factor $T_{j-1}^{-}$.
Note that any wave incident on boundary at $r_{j-1}$ can be further reflected from this boundary, then can be reflected from the boundary at $r_{j}$ and is incident on boundary at $r_{j-1}$ once again.
We write
\begin{equation}
\begin{array}{lcl}
  \tilde{T}_{j-1}^{-} & = &
    \tilde{T}_{j}^{-} T_{j-1}^{-}
    \Bigl(1 + \sum\limits_{m=1}^{+\infty} (R_{j-1}^{-} \tilde{R}_{j}^{+})^{m} \Bigr) =
    \displaystyle\frac{\tilde{T}_{j}^{-} T_{j-1}^{-}} {1 - R_{j-1}^{-} \tilde{R}_{j}^{+}}.
\end{array}
\label{eq.2.2.1.4}
\end{equation}
Here, we use a summarized reflection amplitude $\tilde{R}_{j}^{+}$
(which takes into account transmission of waves through boundary at $r_{j}$,
then after further reflections and transmissions they return back to region with number $j$):
\begin{equation}
\begin{array}{l}
   \vspace{1mm}
   \tilde{R}_{j-1}^{+} =
     R_{j-1}^{+} + T_{j-1}^{+} \tilde{R}_{j}^{+} T_{j-1}^{-}
     \Bigl(1 + \sum\limits_{m=1}^{+\infty} (\tilde{R}_{j}^{+}R_{j-1}^{-})^{m} \Bigr) =
     R_{j-1}^{+} +
     \displaystyle\frac{T_{j-1}^{+} \tilde{R}_{j}^{+} T_{j-1}^{-}} {1 - \tilde{R}_{j}^{+} R_{j-1}^{-}}.
\end{array}
\label{eq.2.2.1.5}
\end{equation}
%
We choose
$\tilde{R}_{N-1}^{+} = R_{N-1}^{+}$ and
$\tilde{T}_{N-1}^{-} = T_{N-1}^{-}$
%
%
and consequently calculate all amplitudes
$\tilde{R}_{N-2}^{+}$ \ldots $\tilde{R}_{N_{\rm cap}}^{+}$,
and $\tilde{T}_{N-2}^{-}$ \ldots $\tilde{T}_{N_{\rm cap}}^{-}$, using recurrent relations above.
We define the summarized amplitude $A_{T}$ of transition through the barrier via all waves transmitted through the potential region with the barrier from $r_{\rm cap}$ to $r_{N-1}$:
$A_{T, {\rm bar}} = \tilde{T}_{N_{\rm cap}}^{-}$.
%

To sum all waves reflected from the boundary at point $r_{j+1}$ and propagating to the right,
we calculate a summarized amplitude of reflection as
\begin{equation}
\begin{array}{lcl}
  \vspace{1mm}
  \tilde{R}_{j+1}^{-} & = &
    R_{j+1}^{-} + T_{j+1}^{-} \tilde{R}_{j}^{-} T_{j+1}^{+}
    \Bigl(1 + \sum\limits_{m=1}^{+\infty} (R_{j+1}^{+} \tilde{R}_{j}^{-})^{m} \Bigr) =
    R_{j+1}^{-} +
    \displaystyle\frac{T_{j+1}^{-} \tilde{R}_{j}^{-} T_{j+1}^{+}} {1 - R_{j+1}^{+} \tilde{R}_{j}^{-}}.
\end{array}
\label{eq.2.2.1.8}
\end{equation}
On such a basis, we define the amplitude of reflection from the potential region with the barrier from $r_{\rm cap}$ to $r_{N-1}$ as
$A_{R, {\rm bar}} = \tilde{R}_{N-1}^{-}$
where $\tilde{R}_{N_{\rm cap}}^{-} = R_{N_{\rm cap}}^{-}$.
%
We find a summarized amplitude $A_{R, {\rm ext}}$ of all waves reflected from the external barrier region
(from the external turning point $r_{\rm tp,ext}$ to $r_{N-1}$)
and propagated outside as
$A_{R, {\rm ext}} = \tilde{R}_{N-1}^{-}$
where $\tilde{R}_{N_{\rm tp,ext}}^{-} = R_{N_{\rm tp,ext}}^{-}$.
%
%
Moreover, we find another summarizing amplitude $A_{R, {\rm tun}}$ of all waves which are reflected just inside the potential region from $r_{\rm cap}$ to the external turning point $r_{\rm tp,ext}$
[i.e. they propagate through the external barrier region without any reflection, tunnels under the barrier, and may propagate to the boundary $r_{\rm cap}$ and further be reflected back from this boundary]
as $A_{R, {\rm tun}} = A_{R, {\rm bar}} - A_{R, {\rm ps}}$.
%
%
We estimate the amplitude of oscillations $A_{\rm osc}$ in region of capture with number $N_{\rm cap}$ as
$A_{\rm osc} (N_{\rm cap}) = (1 - \tilde{R}_{N_{\rm cap}-1}^{-} \tilde{R}_{N_{\rm cap}}^{+})^{-1}$.

In the framework of the MIR formalism, we define the penetrability $T_{\rm bar}$ and the full reflection $R_{\rm bar}$ concerning the barrier (i.e. region from $r_{\rm cap}$ to $r_{N-1}$), the coefficient of reflection $R_{\rm ext}$ of the external part of the barrier (i.e. the region from $r_{\rm tp, ext}$ to $r_{N-1}$), and the coefficient of reflection $R_{\rm tun}$ of the barrier without the external part (i.e. the region from $r_{\rm cap}$ to $r_{\rm tp, ext}$)
as
\begin{equation}
\begin{array}{ccccc}
  T_{\rm bar} \equiv \displaystyle\frac{k_{\rm cap}}{k_{N}}\; \bigl|A_{T, {\rm bar}}\bigr|^{2}, &
  R_{\rm bar} \equiv \bigl|A_{R, {\rm bar}}\bigr|^{2}, &
  R_{\rm ext} \equiv \bigl|A_{R, {\rm ext}}\bigr|^{2}, &
  R_{\rm tun} \equiv \bigl|A_{R, {\rm tun}}\bigr|^{2}.
\end{array}
\label{eq.2.2.2.1}
\end{equation}
We check the property
$T_{\rm bar} + R_{\rm bar} = 1$,
%
%
which indicates whether the MIR method gives proper solutions for the wave function.
%
%
We calculate the summations of amplitudes $\alpha_{j}^{(i)}$ and $\beta_{j}^{(i)}$ for arbitrary region with number $j$:
\begin{equation}
\begin{array}{ll}
  \beta_{j} \equiv
  \displaystyle\sum\limits_{i=1} \beta_{j}^{(i)} =
  \tilde{T}_{j}^{-}\, \Bigl( 1 + \displaystyle\sum\limits_{i=1} (\tilde{R}_{j-1} \tilde{R}_{j}^{+})^{i} \Bigr) =
  \displaystyle\frac{\tilde{T}_{j}^{-}}{1 - \tilde{R}_{j-1} \tilde{R}_{j}^{+}}, &

  \alpha_{j} \equiv
  \displaystyle\sum\limits_{i=1} \alpha_{j}^{(i)} =
  \tilde{R}_{j-1}\, \displaystyle\sum\limits_{i=1} \beta_{j}^{(i)} =
  \displaystyle\frac{\tilde{R}_{j-1} \tilde{T}_{j}^{-}}{1 - \tilde{R}_{j-1} \tilde{R}_{1}^{+}}.
\end{array}
\label{eq.2.2.3.1}
\end{equation}
%
We define \emph{the probability of existence of a compound nucleus via integral over a space region between two internal turning points $r_{\rm int,1}$ and $r_{\rm int,2}$, where point $r_{\rm int,2}$ is middle turning point concerning the barrier for under-barrier energy, or
the coordinate of the maximum of the barrier for above-barrier energy}:
%
%
\begin{equation}
  P_{\rm cn} \equiv
  \displaystyle\int\limits_{r_{\rm int,1}}^{r_{\rm int,2}} |R(r)|^{2}\; r^{2}dr =
  \displaystyle\sum\limits_{j=1}^{n_{\rm int}}
  \Bigl\{
    \bigl( |\alpha_{j}|^{2} + |\beta_{j}|^{2} \bigr)\, \Delta r +
    \displaystyle\frac{\alpha_{j}\beta_{j}^{*}} {2ik_{j}}\,  e^{2ik_{j}r}
      \Bigr|_{r_{j-1}}^{r_{j}} -
    \displaystyle\frac{\alpha_{j}^{*}\beta_{j}} {2ik_{j}}\,  e^{-2ik_{j}r}
      \Bigr|_{r_{j-1}}^{r_{j}}
  \Bigr\}.
\label{eq.2.2.4.3}
\end{equation}
%


There is a traditional definition for the cross-section $\sigma$ of fusion in the $\alpha$ capture, that is based on the penetrabilities $T_{\rm bar, l}$ and probabilities of fusion $P_{l}$.
It is assumed that fusion takes place completely after tunneling of $\alpha$ particle through the barrier;
for example, see Ref.~\cite{Eberhard.1979.PRL}:
%
%
\begin{equation}
\begin{array}{lll}
  \sigma_{\rm fus} (E) = \displaystyle\sum\limits_{l=0}^{+\infty} \sigma_{l}(E), &
  \sigma_{l}(E) = \displaystyle\frac{\pi\hbar^{2}}{2mE}\, (2l+1)\, T_{{\rm bar,} l}(E)\, P_{l},
\end{array}
\label{eq.2.2.5.1}
\end{equation}
where $\sigma_{l}$ is partial cross-section at $l$, and
$E$ is energy of relative motion of the $\alpha$ particle with respect to the nucleus.
To study the compound nucleus, we introduce a new definition for the partial cross section of fusion via the probability of existence of the compound nucleus (\ref{eq.2.2.4.3})
as:
%
%
\begin{equation}
\begin{array}{lll}
  \sigma_{l} = \displaystyle\frac{\pi\hbar^{2}}{2mE}\, (2l+1)\, f_{l}(E)\, P_{\rm cn} (E), &
  f(E) = \displaystyle\frac{k_{\rm cap}}{k_{N}\, |r_{\rm cap} - r_{\rm tp,in, 1}|}.
\end{array}
\label{eq.2.2.5.2}
\end{equation}
To study the compound nucleus living for some period, we apply the idea (\ref{eq.2.1.7.7}) of coefficients of fusion
in the internal nuclear region:
\begin{equation}
\begin{array}{lll}
  T_{j}^{\pm, \rm (new)} \equiv T_{j}^{\pm, \rm (old)} \cdot (1 - p_{j}), &
  R_{j}^{\pm, \rm (new)} \equiv R_{j}^{\pm, \rm (old)} \cdot (1 - p_{j}), &
  0 \le p_{j} \le 1.
\end{array}
\label{eq.2.2.5.3}
\end{equation}
One can see that, with the simple potential (\ref{eq.2.1.1}) for fast complete fusion ($p_{j}=1$), Eq.~(\ref{eq.2.2.5.3}) is transformed to Eq.~(\ref{eq.2.2.5.1}). A similar result is obtained for a potential with a barrier of arbitrary complicated shape.

\section{Analysis
\label{sec.3}}

For analysis
we choose the $^{44}{\rm Ca}$ nucleus.
As shown in Ref.~\cite{Maydanyuk.2015.NPA}, the penetrability is essentially (for the same fixed energy of the incident $\alpha$ particle) dependent on the internal boundary (at $r_{\rm cap}$) of the potential barrier region in calculations.
Therefore, we should impose one additional condition on the determination of the barrier penetrability.
In Ref.~\cite{Maydanyuk.2015.NPA} we proposed a new condition of minimal change of the calculated barrier penetrability at arbitrary variations of $r_{\rm cap}$.
This condition requires that the minimum of the internal potential well should be at this point
(we obtain $r_{\rm cap} = 0.44$~fm at $l=0$ for this nucleus,
in this paper we use parametrization given in Ref.~\cite{Denisov.2005.PHRVA} and
parameters of calculations are 10000 intervals at $r_{\rm max}=70$~fm).
Thus, we use this definition of $r_{\rm cap}$ for further calculations.

\begin{figure}[htbp]
\centerline{\includegraphics[width=86mm]{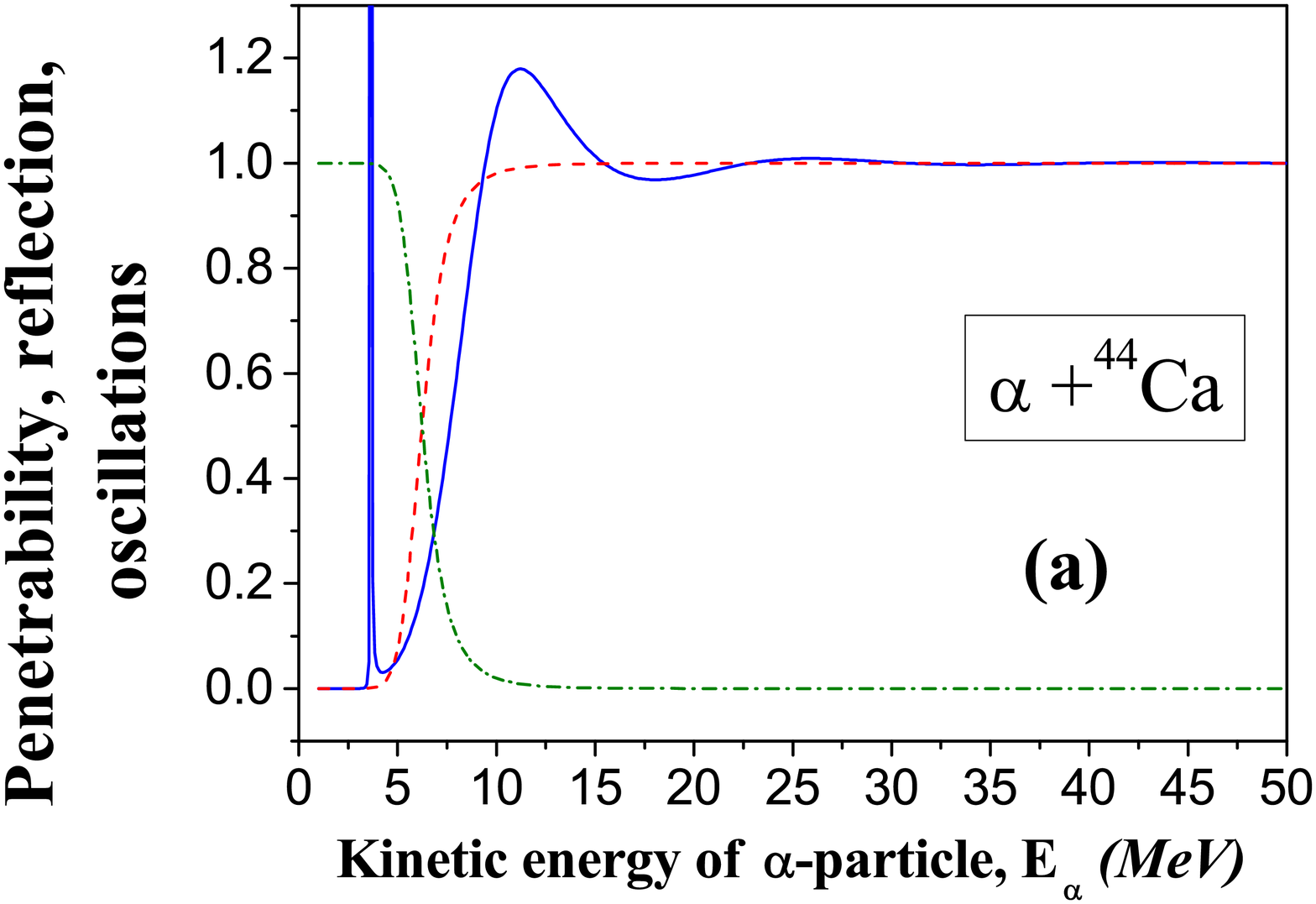}
\hspace{-5mm}\includegraphics[width=86mm]{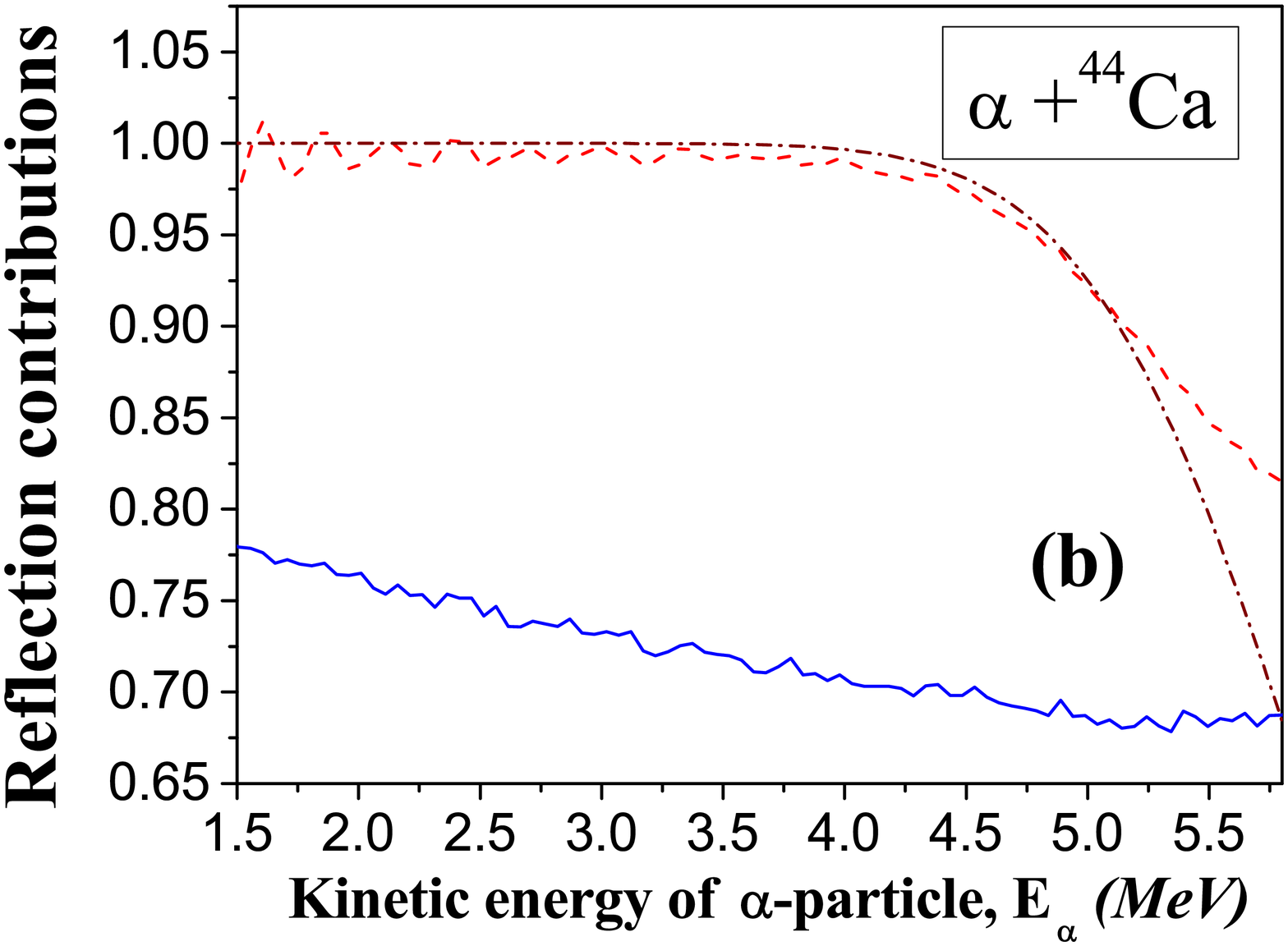}}
\vspace{-4mm}
\caption{\small (Color online)
Panel (a): Penetrability $T_{\rm bar}$ (dashed red line) and reflection $R_{\rm bar}$ (dash-dotted green line) of the barrier defined in Eqs.~(\ref{eq.2.2.2.1}),
modulus of the amplitude of oscillations $|A_{\rm osc}|$ (solid blue line) 
as a function of the energy of the incident $\alpha$ particle for the reaction of $\alpha + ^{44}{\rm Ca}$ at $l=0$
(test of $T_{\rm bar} + R_{\rm bar} =1$ is fulfilled up to the first 14 digits for
all energy levels considered for all results in this paper).
One can see maxima (the first one is in under-barrier-energy region) of the amplitude of oscillation.
%
Panel (b): Coefficients of reflection $R_{\rm ext}$ and $R_{\rm tun}$ defined in Eqs.~(\ref{eq.2.2.2.1})
as a function of the energy $E_{\alpha}$ of the incident $\alpha$ particle for the reaction of $\alpha + ^{44}{\rm Ca}$ at $l=0$.
Here, the solid blue line is for the coefficient of reflection $R_{\rm ext}$,
the dashed red line is for the coefficient of reflection $R_{\rm tun}$, and
the dash-dotted brown line is for the full reflection $R_{\rm bar}$.
One can see that, up to good accuracy, the full reflection is determined by waves
propagating via the stage of the compound nucleus formation and its disintegration (i.e. by $R_{\rm tun}$).
However, the potential scattering is not small and is close enough to full reflection inside the full analyzed energy region.
This result contradicts with a popular point of view that the reflection (in capture and decay nuclear tasks) is formed just by the internal tunneling processes inside the barrier.
One can propose formula $P_{\rm ref, ps} \approx 0.75 \cdot P_{\rm ref}$ at $1.5~{\rm MeV} < E_{\alpha} <5~{\rm MeV}$ for quick approximated estimations.
\label{fig.3.1}}
\end{figure}

In Fig.~\ref{fig.3.1}~(a) we show the penetrability, reflection and amplitude of oscillation
in dependence on the energy of the incident $\alpha$ particle at $l=0$.
One can see that the modulus of the amplitude of oscillations has sharp maxima while the penetrability and reflections are monotonous functions.
This result is the first indication on the existence of maximally stable states of the compound nucleus that lives some periods at definite energies of the incident $\alpha$ particle (where one level is under the barrier energy). Note that the penetrability does not provide any information about such states.
For completeness, we add our calculations of the coefficients of reflection $R_{\rm ext}$ and $R_{\rm tun}$
in Fig.~\ref{fig.3.1}~(b).
%
%

\begin{figure}[htbp]
\centerline{\includegraphics[width=86mm]{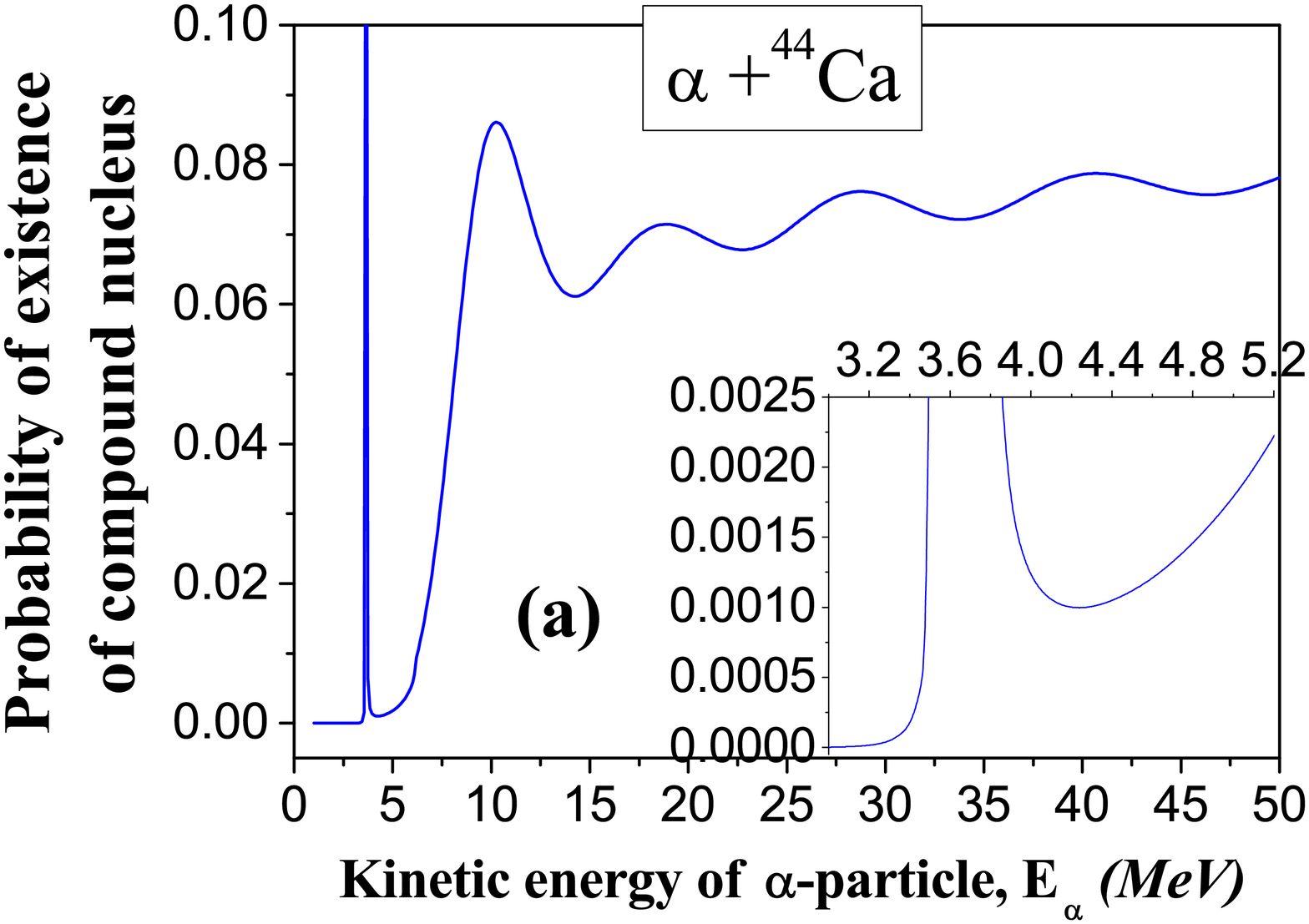}
\hspace{-5mm}\includegraphics[width=86mm]{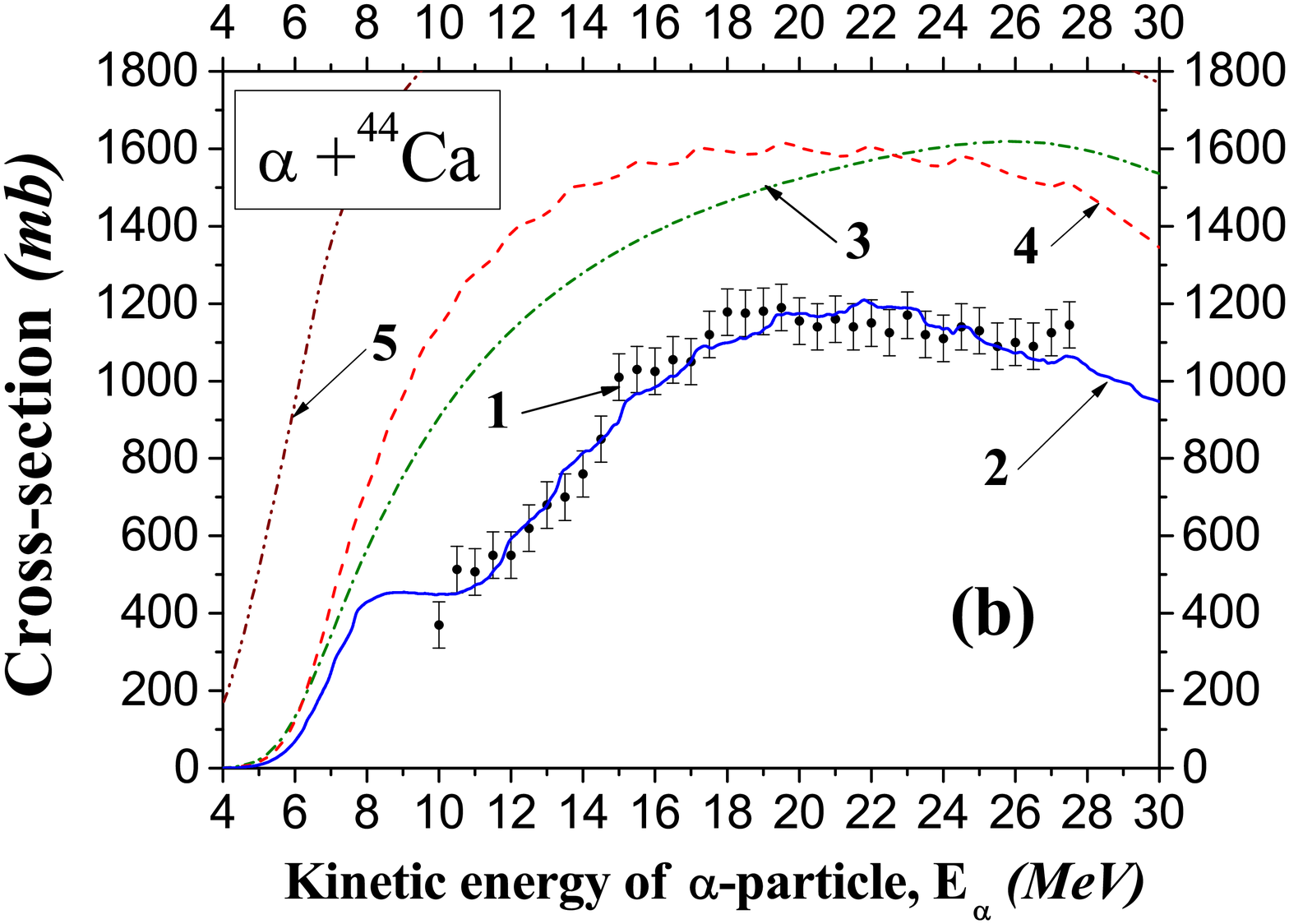}}
\vspace{-4mm}
\caption{\small (Color online)
Panel (a): The probability $P_{\rm cn}$ of existence of the compound nucleus defined in Eqs.~(\ref{eq.2.2.4.3}) in dependence on the energy of the incident $\alpha$ particle
for the reaction of $\alpha + ^{44}{\rm Ca}$ at $l=0$.
The function $P_{\rm cn}$ has an oscillator behavior, with the clear presence of five maxima (here the first maximum is in under-barrier-energy region).
For under-barrier energies above this first quasibound energy level there is one minimum (at $E_{\rm min}=4.24$~MeV, $P_{\rm cn}=0.000986$), indicating that the nucleus becomes more transparent for penetration of the $\alpha$ particle.
At energies closer to zero, the probability of formation of the compound nucleus decreases quickly to zero.
One can see a stable picture of $P_{\rm cn}$ near this first quasibound energy level.
Note that penetrability and reflection coefficients $T_{\rm bar}$ and $R_{\rm bar}$ have no any picks near this energy (see Fig.~\ref{fig.3.1}).
%
Panel (b): Complete cross section of fusion with the included probabilities of fusion in dependence on the energy of the incident $\alpha$ particle for the reaction $\alpha + ^{44}{\rm Ca}$
(parameters of calculations: 1000 intervals at $r_{\rm max}=70$~fm).
Here, the data labeled 1 are the experimental data extracted from Ref.~\cite{Eberhard.1979.PRL},
the solid blue line 2 is a cross section defined by Eqs.~(\ref{eq.2.2.5.2}) and (\ref{eq.2.2.5.3}) with included probabilities of fusion [their values are given in Table~\ref{table.3.2.1},
the accuracy of agreement with experimental data is $\varepsilon_{1}=0.033308$,
$\varepsilon_{1}$ is defined in \cite{Maydanyuk.2015.NPA}],
the dash-dotted green line 3 is cross section in the old definition~(\ref{eq.2.2.5.1}), where penetrabilities are calculated by the MIR approach,
the dashed red line 4 is a cross section defined in Eqs.~(\ref{eq.2.2.5.2}) and (\ref{eq.2.2.5.3}) without the coefficients of fusion,
the dash-double dotted brown line 5 is a cross section in the old definition~(\ref{eq.2.2.5.1}), where the penetrabilities are calculated by using the WKB approximation.
\label{fig.3.2}}
\end{figure}
\begin{figure}[htbp]
\centerline{\includegraphics[width=86mm]{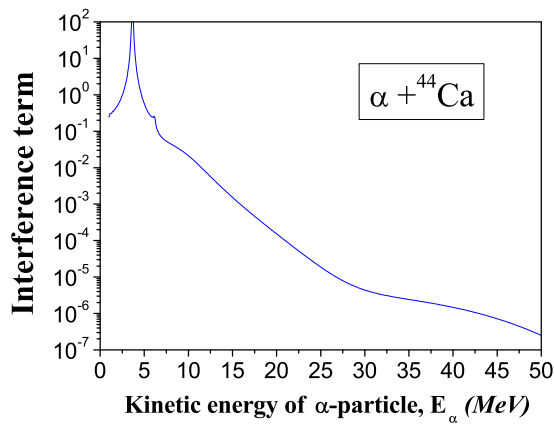}}
\vspace{-5mm}
\caption{\small (Color online)
Interference term $P_{\rm interf}$
in dependence on the energy of the incident $\alpha$ particle for the reaction of $\alpha + ^{44}{\rm Ca}$ at $l=0$.
\label{fig.3.3}}
\end{figure}

In Fig.~\ref{fig.3.2}~(a) we present our calculations for the probability $P_{\rm cn}$ of the existence of the compound nucleus.
One can see the presence of clear maxima in that dependence of the function $P_{\rm cn}$ on energy (here the first maximum of the function $P_{\rm cn}$ is in the under-barrier-energy region).
These maxima should be interpreted as indication on the most stable (i.e., lived for the most long time) states of the formed compound nucleus.
Note that there are no any maxima in dependence of the penetrability $T_{\rm bar}$ on energy in this energy region (see Fig.~\ref{fig.3.1}).
These dependencies have been used in the basis of the traditional calculations of the cross section of the capture of the $\alpha$ particle by a nucleus (for example, see Eqs.~(1) and (2) in Ref.~\cite{Eberhard.1979.PRL}).
We call such states of the compound nucleus (and the corresponding energy values) as ``quasibound states''.
A unified description of the presence of these states of the compound nucleus, a prediction of the corresponding energy values and monotonic penetrabilities are advances of the Method of the multiple internal reflections.
The clearest understanding of the presence of these quasibound states of the compound nucleus at monotonic penetrability of the barrier can be obtained from the simplest $\alpha$-capture picture studied above.
In particular, here one can see that the modulus $|A_{\rm osc}|$ in Eq.~(\ref{eq.2.1.3.11}) [and corresponding sums of amplitudes in Eqs.~(\ref{eq.2.1.3.11})] has a clear maximums which can be larger unity.
The incident wave in the external region is normalized on unity, and all these tests confirm this formalism and calculations with high accuracy.
Note that accurate information about the quasibound states for above-barrier energies cannot be extracted from the interference term (there is only one clear maximum at $E_{\alpha}=3.651$~MeV in the interference term at $l=0$ in Fig.~\ref{fig.3.3} in comparison with five peaks of $P_{\rm cn}$ in Fig.~\ref{fig.3.1}~(a),
the corresponding five energy values are shown in
Table~\ref{table.3.2.2})%
\footnote{We do not analyze possible very small variations of the interference term (like small peak at $E_{\alpha}=6.106$~MeV in Fig.~\ref{fig.3.3}) in this paper,
these are caused by the numeric calculations and are not connected with the quasibound states.}.

To estimate the fusion in the studied reaction,
let us understand how closely the formula~(\ref{eq.2.2.5.2}) provides the cross section in a comparison with the old definition~(\ref{eq.2.2.5.1}).
In this paper we use the same fusion probabilities $p_{i}^{\rm (int)}$ inside the region from $r_{\rm int,1}$ to $r_{\rm cap}$, and the same fusion probabilities $p_{i}^{\rm (ext)}$ inside the region from $r_{\rm cap}$ to $r_{\rm int,2}$.
For fast complete fusion we have $p_{i}^{\rm (int)}=1$ and $p_{i}^{\rm (out)}=0$.
Such a calculated cross section and the old result are presented in Fig.~\ref{fig.3.2}~(b) (see dashed red line 4 line and dash-dotted green line 3 line, respectively).
One can see that new calculations are enough close to the previous results
(they have similar shapes and have no resonances),
so the new definition~(\ref{eq.2.2.5.2}) is applicable for analysis of the fusion in this reaction.
On such a basis, now we investigate the possibility of evolution of the compound nucleus and its disintegration (where the fusion probabilities are not equal to unity),
and estimate the fusion via variation of the fusion probabilities.

The result of such an analysis is presented in Fig.~\ref{fig.3.2}~(b) by the blue solid line 2.
One can see that inclusion of the fusion probabilities allows us to increase agreement between theory and experimental data essentially.
In Table~\ref{table.3.2.1} we present the fusion probabilities.
\begin{table}
\begin{center}
\begin{tabular}{|c|c|c|c|c|c|c|c|c|c|c|c|c|c|c|c|c|c|c|c|c|c|} \hline
  & $l=0$ & $l=1$ & $l=2$ & $l=3$ & $l=4$ & $l=5$ & $l=6$ & $l=7$ & $l=8$ & $l=9$ &
           $l=10$ & $l=11$ & $l=12$ & $l=13$ & $l=14$ & $l=15$ & $l=16$
    \\ \hline
 $p_{l}^{\rm (int)}$ & 0 & 0 & 0 & 0 & 0 & 1.00 & 1.00 & 0 & 1.00 & 1.00 & 0 & 0 & 1.00 & 0.125 & 1.00 & 0 & 0
   \\ \hline
 $p_{l}^{\rm (ext)}$ & 1.00 & 1.00 & 1.00 & 0 & 1.00 & 1.00 & 0 & 0.93 & 0 & 0 & 0.875 & 0.94 & 0 & 0.875 & 0 & 0.625 & 1.00 \\ \hline
\end{tabular}
\end{center}
\vspace{-5mm}
\caption{Fusion probabilities for the cross section
presented in Fig.~\ref{fig.3.2}~(b) (see solid blue line 2 in that figure).}
\label{table.3.2.1}
\end{table}
One can see that some fusion probabilities are not equal to unity, that indicates that complete fusion in those channels is not fast.
Thus, in such channels further propagation of waves without fusion (or with partial fusion) takes place inside the internal nuclear region after tunneling, i.e., the compound nucleus is formed and it evolves for some time. For such channels, we find energy values for quasibound states where the compound nucleus is the most stable.
In Table~\ref{table.3.2.2} we calculate the quasibound energy values for the reaction $\alpha + ^{44}{\rm Ca}$.
%
%
\begin{table}
\begin{center}
\begin{tabular}{|c|c|c|c|c|c|c|c|c|c|c|} \hline
  $l=0$ &  $l=1$ &  $l=2$ &  $l=3$ &  $l=4$ &  $l=5$ &  $l=6$ &  $l=7$ & $l=14$ \\ \hline
  3.651 &  6.597 & 3.356 &  6.499 &  9.543 &  6.008 &  9.248 & 12.488 & 21.621 \\
 10.328 & 14.649 & 10.034 & 14.354 & 18.380 & 13.569 & 18.184 & 23.094 & 33.993 \\
 18.969 & 23.486 & 18.969 & 23.683 & 28.985 & 23.683 & 29.084 & 34.975 &  - \\
 28.789 & 34.681 & 28.887 & 35.270 & 41.358 & 35.074 & 41.653 & 49.018 &  - \\
 40.867 & 47.446 & 41.064 & 48.036 &   -    & 48.527 &   -    &   -    &  - \\ \hline
\end{tabular}
\end{center}
\vspace{-6mm}
\caption{Predicted energy values (in MeV) for the quasibound states of the compound nucleus
formed in the capture reaction of $\alpha + ^{44}{\rm Ca}$ up to 50~MeV for the first some $l$
(parameters of calculations: 10000 intervals at $r_{\rm max}=70$~fm).
%
}
\label{table.3.2.2}
\end{table}

\section{Comparison with theory of quasistationary states with complex energies
\label{sec.3.2}}

Today there is a theory of quasistationary states with complex energies~\cite{Baz.1971.book} which allows us to determine the energies of quasistationary states in decay tasks.
These quasistationary states correspond to poles of the $S$-matrix with complex energies (for example, see Refs.~\cite{Vertse.1982.CPC,gamow_functions.all}).
This theory is also applied to analyze resonant states in scattering and could be used to calculate energies for the capture processes.
Our comparative analysis shows that this theory gives the quasistationary states, however, these states are not the states given by our approach%
\footnote{The theory of quasistationary states with complex energies~\cite{Baz.1971.book} provides quasistationary states in order to describe nonstability (i.e. non-stationarity) of formed nuclear system in scattering, also to describe nuclear system in decay, capture, etc.
We introduce a new term ``quasibound'' for the states of the most probable existence of the compound nucleus because our approach is realized at real energies of the incident $\alpha$ particle,
as a formal middle case between the bound and nonbound states in standard quantum mechanics (with real energies).}.
We choose the formalism in Ref.~\cite{Vertse.1982.CPC} for analysis of the theory pointed out above.
As we see, the clear difference between the theory of quasistationary states with complex energies and our approach can be obtained from an analysis of two different aspects,
such as the determination of the cross-sections of the $\alpha$-capture and determination of the states (and corresponding energy levels) for the $\alpha$--nucleus interactions.

\subsection{Determination of cross sections of $\alpha$-capture
\label{sec.3.2.1}}

In the first aspect pointed out above, we analyze the applicability of the compared approaches for determination of the cross sections of the $\alpha$-capture.
According to the modern models of the $\alpha$ capture (see Refs.~\cite{Denisov.2009.ADNDT,Maydanyuk.2015.NPA,Denisov.2005.PHRVA}, also Refs.~\cite{Glas.1975.NPA,Glas.1974.PRC}), the cross section of the $\alpha$ capture is determined on the basis of the penetrability of the barrier.
In particular, an accurate determination of the penetrability is a crucial point for a successful calculation of the cross-section.
Our approach provides the unified formalism to calculate the penetrabilities and probabilities of the formation of the compound nucleus.
However, the theory of quasistationary states with complex energies (for example, see approach~\cite{Vertse.1982.CPC} for details) does not determine these characteristics.
Thus, without further modifications, it cannot be used to calculate the penetrabilities and cross sections of $\alpha$ capture in frameworks of the modern models of $\alpha$-capture.

The penetrability is changed varying the space localization of the capture of the $\alpha$ particle by nucleus (see Ref.~\cite{Maydanyuk.2015.NPA} for details and demonstrations; also Refs.~\cite{Maydanyuk.MIR.all}).
This property follows directly from the definition of the penetrability in quantum mechanics. Importantly, this change of the penetrability is not small for the majority of nuclear processes (we estimated it could be up to four times for capture of the $\alpha$ particle by the $^{40,44}{\rm Ca}$ nuclei; for the inverse nuclear processes, such as $\alpha$ decays, this change is essentially larger).
However, the theory of quasistationary states with complex energies ignores this point (so, it is simpler and can be faster).
Our formalism resolves this problem with very good accuracy (we demonstrated this point in Ref.~\cite{Maydanyuk.2015.NPA} in details with many demonstrations).

\subsection{Determination of quantum states of $\alpha$--nucleus elastic scattering
\label{sec.3.2.2}}

In the second aspect, we analyze the applicability of the compared approaches to determine the quantum states in the $\alpha$-capture and scattering.
As we can see, there is a clear difference between the quantum states given by the theory of quasistationary states with complex energy and our approach.
This is shown from the analysis of the elastic scattering of $\alpha + ^{44}{\rm Ca}$ at $l=0$ by both methods, if to use the simplest potential of form (\ref{eq.2.1.1}).
However, the formalism~\cite{Vertse.1982.CPC} cannot directly describe such a reaction, because we should modify the asymptotic boundary condition (5)--(6) of that paper in the form of Eqs.~(\ref{eq.2.1.1}) (at $r_{1} < r$) and choose real values of energy.
After such a modification, unfortunately, analyzing poles or zeros of the $S$-matrix does not give anything, because the modulus of the $S$-matrix equals unity: $|S|=1$ (i.e., there is no zero or pole of the $S$-matrix).
One can see this from the exact analytical $S$-matrix, which is easily obtained by our formalism ($R_{0}=1$):
\begin{equation}
  S = A_{R} =
  \alpha_{2}^{(1)} + \displaystyle\sum\limits_{i=2} \alpha_{2}^{(i)} =
  R_{1}^{-} + A_{\rm osc}\, T_{1}^{-} R_{0} T_{1}^{+}.
\label{eq.3.2.1}
\end{equation}
In contrary, we present calculations of the probability of existence of the compound nucleus in Fig.~\ref{fig.2.1.1} by our approach, where one can clearly see maximums.
We calculate real energies (and the wave functions) corresponding to the maximal probabilities.
Note that we use the same boundary conditions imposed on the radial wave function at zero ($\chi(0)=0$) and the same normalization of the incident wave in the asymptotic [$\chi_{\rm inc}(r) = e^{-ik_{2}r}$] in both approaches.
Moreover, the difference between the two approaches exists as well, if to use the realistic $\alpha$-nucleus potential instead of the simple potential~(\ref{eq.2.1.1}).

\begin{figure}[htbp]
\centerline{\includegraphics[width=86mm]{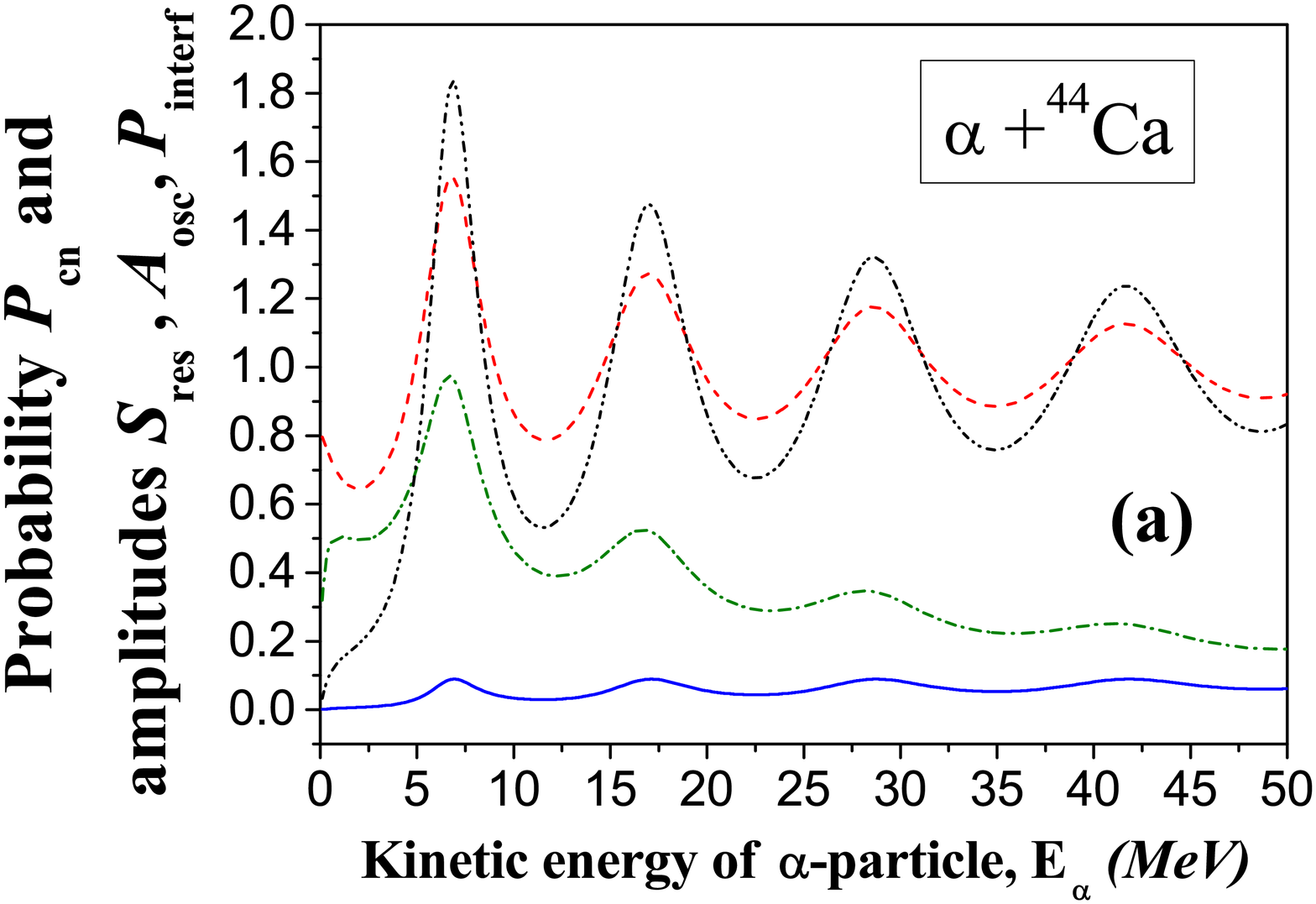}
\hspace{-5mm}\includegraphics[width=86mm]{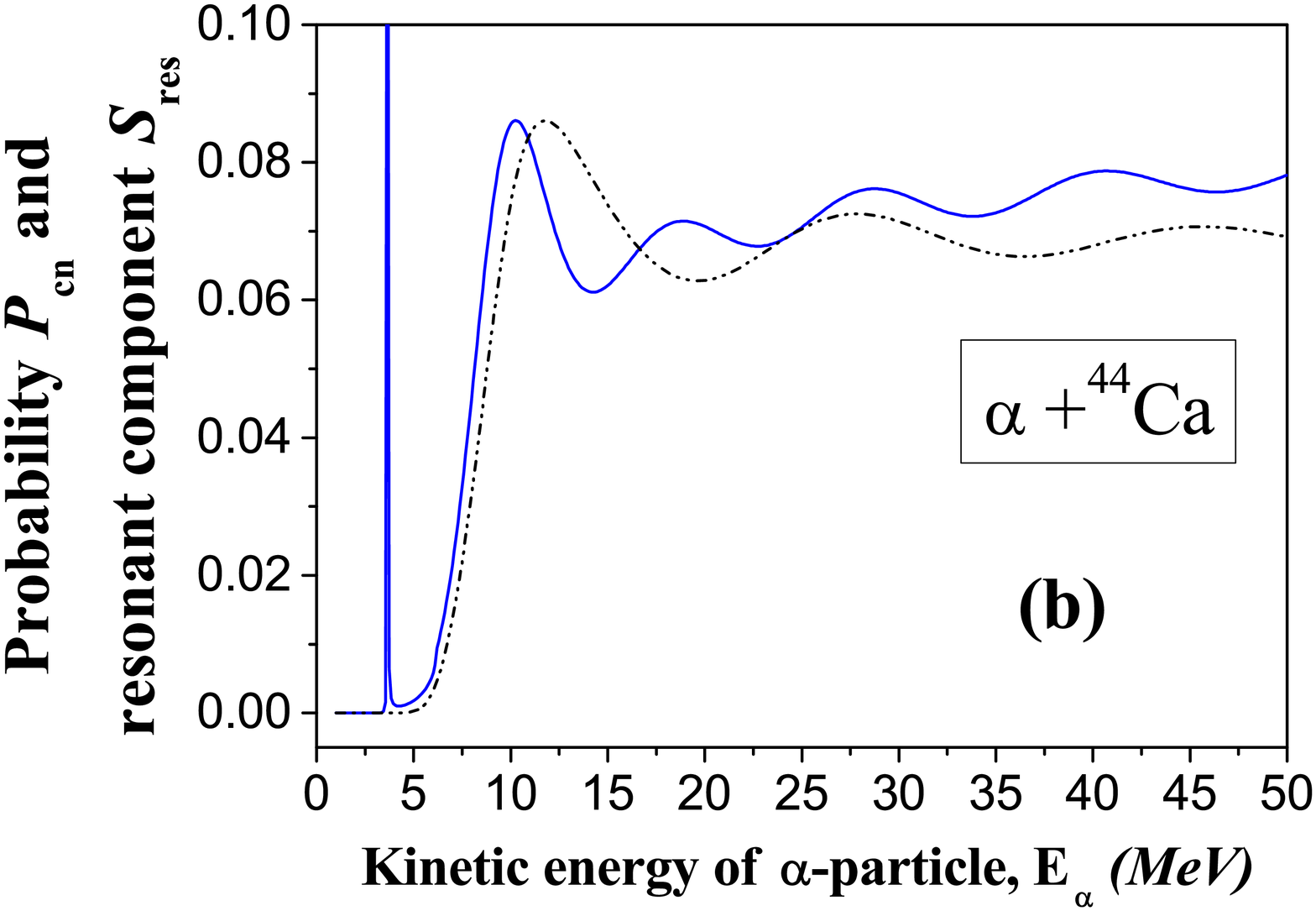}}
\vspace{-4mm}
\caption{\small (Color online)
Panel (a): Modulus of resonant term of the $S$-matrix $S_{\rm res}$ (black dash-double dotted line) in comparison with
the probability of existence of the compound nucleus $P_{\rm cn}$ (blue solid line) defined by Eq.~(\ref{eq.2.1.5.4}),
modulus of the amplitude of oscillations $A_{\rm osc}$ (red dashed line) defined in Eq.~(\ref{eq.2.1.3.11}) and
interference term $P_{\rm interf}$ (green dash-dotted line) defined in Eq.~(\ref{eq.2.1.6.2})
in dependence on the energy of the incident $\alpha$ particle for the reaction $\alpha + ^{44}{\rm Ca}$ at $l=0$ for the simple potential (\ref{eq.2.1.1})
(parameters $V_{1}$ and $r_{1}$ are taken concerning the depth of well and coordinate of the maximum of the realistic radial barrier at parametrization~\cite{Denisov.2005.PHRVA} for this reaction:
we obtain $V_{1}=-23.73$~MeV and $r_{1}=8.935$~fm).
Values of energies for the maxima for the analyzed coefficients are presented in Table~\ref{table.3.2.1}.
Panel (b): Normalized modulus of the resonant term of the $S$-matrix $S_{\rm res}$ (black dash-double dotted line) in comparison with
the probability of existence of the compound nucleus $P_{\rm cn}$ (blue solid line) 
in dependence on the energy of the incident $\alpha$ particle for reaction $\alpha + ^{44}{\rm Ca}$ at $l=0$ for the realistic potential.
In calculations we define the resonant term as $S_{\rm res} = \alpha_{\rm tp,out} - A_{R, {\rm ext}}$, $\alpha_{\rm tp, out}$ is the amplitude of the wave function close to the external turning point.
One can see that the maxima for the presented lines differ, which indicates different states characterized by the probability $P_{\rm cn}$
and resonant term $S_{\rm res}$.
%
\label{fig.3.2.1}}
\end{figure}

More useful information could be obtained if we include resonant and potential terms of the $S$-matrix in the analysis of the elastic scattering above.
The MIR approach determines these components clearly.
For the potential~(\ref{eq.2.1.1}) we have
\begin{equation}
\begin{array}{lll}
  S_{\rm res} =
    \displaystyle\sum\limits_{i=2} \alpha_{2}^{(i)} =
    A_{\rm osc}\, T_{1}^{-} R_{0} T_{1}^{+}, &
  S_{\rm pot} = \alpha_{2}^{(1)} = R_{1}^{-}.
\end{array}
\label{eq.3.2.2}
\end{equation}
The most probable formation of the compound nucleus can be characterized by maxima of the resonant component.
Thus, to compare the $S$-matrix analysis (the formalism~\cite{Vertse.1982.CPC} is based on it) and the approaches MIR (in study of the compound nucleus), we have to compare maxima of the resonant term $S_{\rm res}$ and the probability $P_{\rm cn}$.
We have
\begin{equation}
\begin{array}{lll}
  |S_{\rm res}|^{2} =
    |A_{\rm osc}|^{2} \Bigl| \displaystyle\frac{4kk_{1}}{(k+k_{1})^{2}} \Bigr|^{2}, &
  P_{\rm cn} =
  |A_{\rm osc}|^{2} \displaystyle\frac{4k}{k+k_{1}} \Bigl( r_{1} - \displaystyle\frac{\sin 2k_{1}r_{1}}{2k_{1}} \Bigr) =
  |A_{\rm osc}|^{2} P_{\rm loc} \displaystyle\frac{2k_{1}}{k+k_{1}}
\end{array}
\label{eq.3.2.2}
\end{equation}
or
\begin{equation}
  P_{\rm cn} = |S_{\rm res}|^{2} \cdot P_{\rm loc}\, \displaystyle\frac{(k+k_{1})^{3}}{8k_{1}k^{2}}.
\label{eq.3.2.3}
\end{equation}
%
From these formulas, we obtain different maxima of the existence probabilities of the our compound nucleus and the resonant component of the $S$-matrix.
In Fig.~\ref{fig.3.2.1}~(a) we present our calculations with such characteristics 
for reaction $\alpha + ^{44}{\rm Ca}$ at $l=0$ for the simple potential (\ref{eq.2.1.1}).
One can see that maxima for the resonant term $S_{\rm res}$ and the probability $P_{\rm cn}$ are close, but different (see Table~\ref{table.3.2.1}).
\begin{table}
\begin{center}
\begin{tabular}{|c|c|c|c|c|c|c|c|c|c|c|} \hline
                 & $A_{\rm osc}$ & $P_{\rm cn}$ & $P_{\rm interf}$ & $S_{\rm res}$ \\ \hline
  $E_{\rm max, 1}$ (MeV) &   -    &   -              &  0.934 & - \\
  $E_{\rm max, 2}$ (MeV) &  6.775 &  6.942 &  6.608 &  6.775 \\
  $E_{\rm max, 3}$ (MeV) & 16.955 & 17.122 & 16.622 & 16.955 \\
  $E_{\rm max, 4}$ (MeV) & 28.638 & 28.805 & 28.137 & 28.638 \\
  $E_{\rm max, 5}$ (MeV) & 41.655 & 41.822 & 40.987 & 41.655 \\ \hline
\end{tabular}
\end{center}
\caption{Energies for maxima of modulus of amplitude of oscillations $A_{\rm osc}$, probability of existence of compound nucleus $P_{\rm cn}$, interference term $P_{\rm interf}$, and modulus of resonant term of $S$-matrix, $S_{\rm res}$, for reaction $\alpha + ^{44}{\rm Ca}$ at $l=0$ for the simple potential (\ref{eq.2.1.1}).
One can see that these energies for $S_{\rm res}$ are coincident with energies for maxima for $A_{\rm osc}$ but differ from energies for maxima for $P_{\rm cn}$.
It is interesting to note that maximal probabilities at such energies are almost the same: $P_{\rm cn, max}=0.0901$.
%
%
%
}
\label{table.3.2.1}
\end{table}
Thus, we conclude that the approach based on analysis of the resonant component of the $S$-matrix and our approach determining the probability $P_{\rm cn}$ describe different states of the compound nucleus.
In particular, energy shifts between maxima of $S_{\rm res}$ and $P_{\rm cn}$ can be determined by factor $P_{\rm loc}\, (k+k_{1})^{3} / (8k_{1}k^{2})$.
Note that the formalism~\cite{Vertse.1982.CPC} does not give any explanation of the relationship between the resonant scattering and internal processes inside the well, while we provide an accurate unified formalism describing them.
This is the advance of our approach, which has no alternative methods in quantum physics, at present.

In Fig.~\ref{fig.3.2.1} (b) we present our calculations for the normalized modulus of the resonant term of the $S$-matrix, $S_{\rm res}$, in comparison with
the probability of existence of the compound nucleus $P_{\rm cn}$ 
for the reaction $\alpha + ^{44}{\rm Ca}$ at $l=0$ for the realistic potential.
Once again, we see that maxima are different.
Note that, in contrast with the WKB approach, in the fully quantum analysis the potential and resonant scattering are already dependent on an additional independent parameter defining the external boundary of the potential region with barrier (we chose it as the external turning point).


\subsection{Determination of quantum states of $\alpha$--nucleus inelastic scattering and $\alpha$ capture
\label{sec.3.2.3}}

\begin{figure}[htbp]
\centerline{\includegraphics[width=86mm]{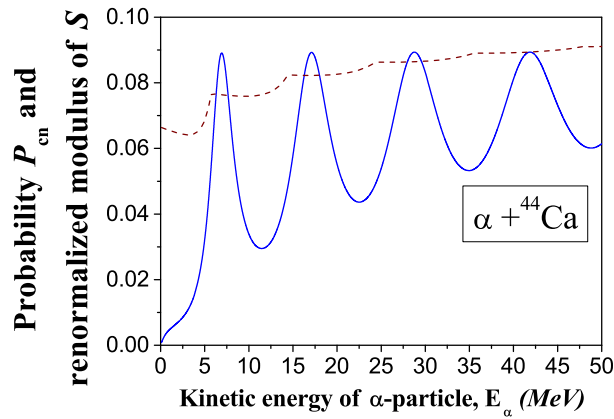}}
\vspace{-4mm}
\caption{\small (Color online)
Renormalized modulus of the $S$-matrix at complex energy (brown dash-double dotted line) compared with the probability of existence of the compound nucleus $P_{\rm cn}$ (blue solid line) presented in Fig.~(\ref{fig.3.2.1}),
in dependence on the real part of energy of the incident $\alpha$ particle for the reaction $\alpha + ^{44}{\rm Ca}$ at $l=0$ for the simple potential (\ref{eq.2.1.1})
(parameters $V_{1}$ and $r_{1}$ of the potential are taken as in calculations of Fig.~\ref{fig.2.1.1}).
One can see that the modulus of the $S$-matrix has no zero in the studied energy region, that indicates on absence complete capture in frameworks of the formalism~\cite{Vertse.1982.CPC}.
Minima of the modulus of the $S$-matrix do not correspond to maxima of the probability $P_{\rm cn}$.
Thus, we have the different resonating energies calculated by our approach and generalization of the S-matrix approach describing states of the most probable existence of the compound nucleus.
%
\label{fig.3.2.2}}
\end{figure}

Inclusion of the complex energies in our analysis allows us to add inelastic processes into our task.
Here, the Multiple internal reflection method can be easily generalized to the calculations with the complex energy of the incident $\alpha$ particle, because the formalism uses complex values for wave numbers $k_{i}$, amplitudes of wave function $\alpha_{i}$ and $\beta_{i}$, amplitudes $T_{i}^{\pm}$ and $R_{i}^{\pm}$, etc. in each potential region.
However, direct application of the formalism~\cite{Vertse.1982.CPC} to the studied reaction with the simplified potential~(\ref{eq.2.1.1}) does not give any solution in real energy region up to 50~MeV, that means there is no complete capture according to this approach.
In Fig.~\ref{fig.3.2.2} we present our calculations of the modulus of the $S$-matrix with complex energy.
Here, one can see that this function has no zero in the studied energy region.%
\footnote{Zero of the $S$-matrix corresponds to the boundary condition of zero outgoing wave in the asymptotic limit.
This is along the main idea of the formalism~\cite{Vertse.1982.CPC} adapted for the capture process.}
One can generalize the formalism~\cite{Vertse.1982.CPC} and supposing that minima of the modulus of the $S$-matrix correspond to states of the most probable formation of the compound nucleus.
But, as one can see from this figure, these minima do not coincide with maxima of the probability of existence of the compound nucleus $P_{\rm cn}$, calculated by our approach above.
Such a picture clearly shows that this modification of the formalism~\cite{Vertse.1982.CPC} and our approach describe different states of the compound nucleus.

In next Fig.~\ref{fig.3.2.3} we present our calculations for the modulus of the $S$-matrix and the corresponding $\Gamma$-width for the realistic $\alpha$-nucleus potential at complex energy of the incident $\alpha$ particle
(
we chose real energy region up to 7~MeV).
\begin{figure}[htbp]
\centerline{\includegraphics[width=86mm]{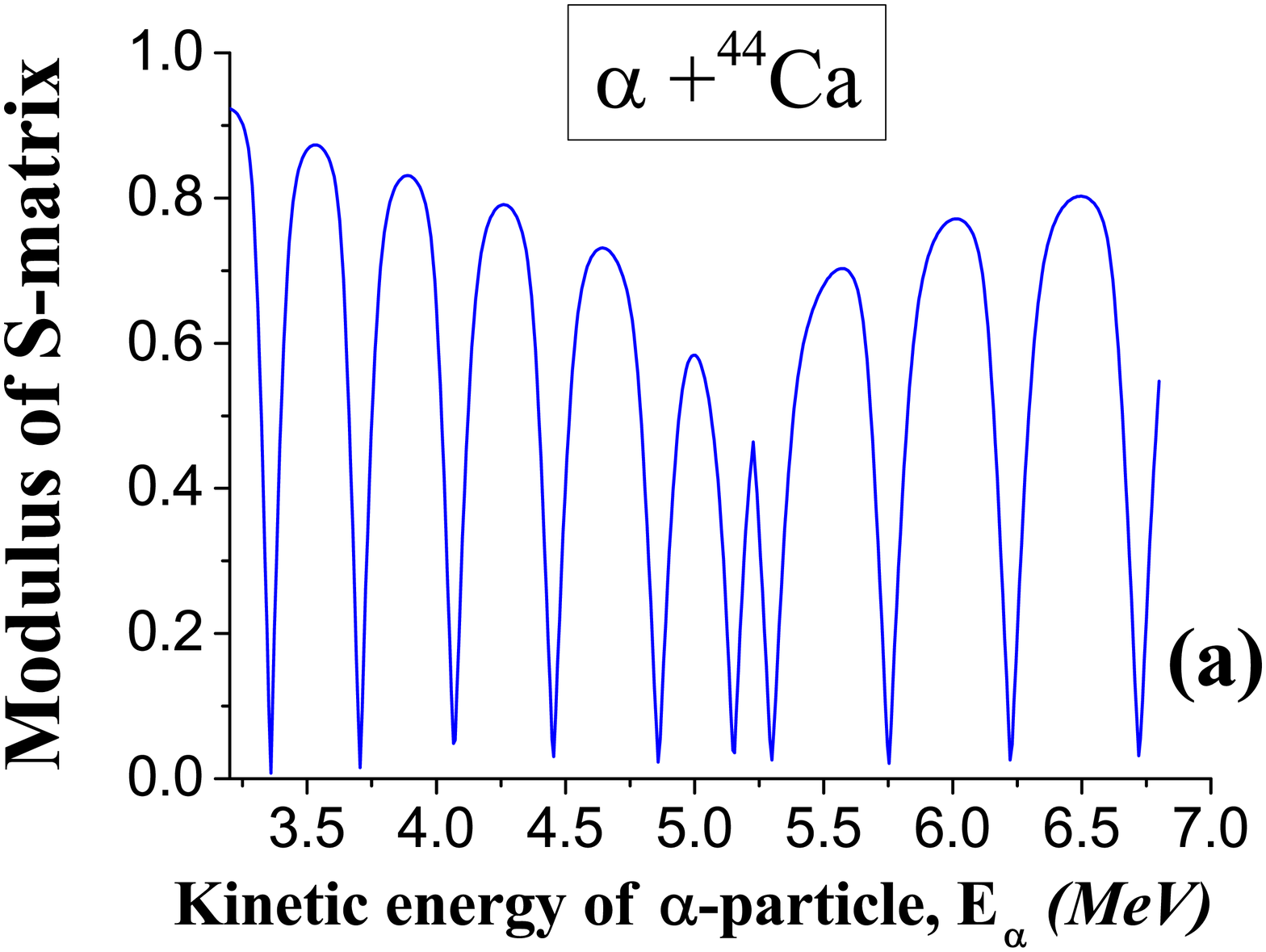}
\hspace{-5mm}\includegraphics[width=86mm]{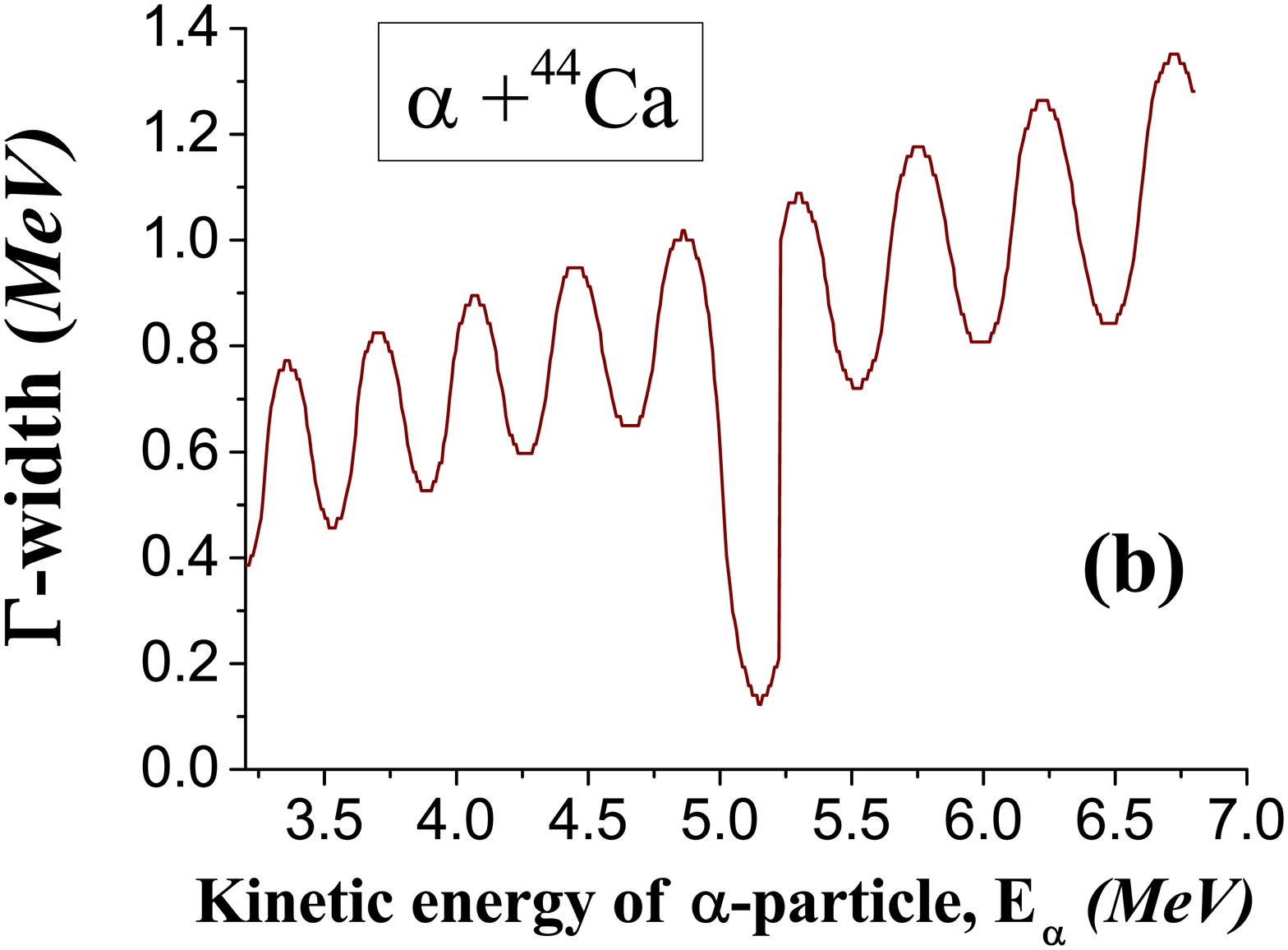}}
\vspace{-3mm}
\caption{\small (Color online)
Panel (a):
Modulus of the $S$-matrix at complex energy 
in dependence on the real part of energy of the incident $\alpha$ particle for the reaction $\alpha + ^{44}{\rm Ca}$ at $l=0$ for the realistic potential.
Panel (b):
The $\Gamma$-width in dependence on the real part of the energy of the incident $\alpha$ particle
corresponding to the calculated $S$ matrix shown in panel (a).
\label{fig.3.2.3}}
\end{figure}
One can see that, inside the analyzed energy region the modulus of the $S$-matrix has 10 minima.
According to logic and the main positions of the theory of quasistationary states with complex energies, these minima are very close to zero and correspond to the most probable states of the possible $\alpha$-capture.
Upon comparing this result with results given in Fig.~\ref{fig.3.2} (a) for the calculated probability of existence of the compound nucleus $P_{\rm cn}$ (we have five maxima of that function in the energy region up to 50~MeV, see Table~\ref{table.3.2.2}), we conclude that these states calculated at minima (zero) of the $S$-matrix at complex energy of the incident $\alpha$ particle are essentially different from states given by
maxima of the probability of existence of the compound nucleus calculated by our approach.

\section{Conclusions
\label{sec.conclusions}}

In this paper we study capture of the $\alpha$-particles with nuclei by the improved MIR method.
We discover new most stable states (called as quasibound states) of a compound nucleus formed in this reaction.
With a simple example (see Fig.~\ref{fig.2.1.1} and explanations in caption to that figure), we explain the absence of these states in traditional calculations of $\alpha$-capture cross-sections as following:
The penetrabilities of barrier have monotonous dependencies on the energy of the $\alpha$ particle (see Fig.~2 in Ref.~\cite{Maydanyuk.2015.NPA}, for details).
Based on such monotonous penetrabilities, full cross sections of the $\alpha$ capture have no peaks (see Fig.~3 in Ref.~\cite{Maydanyuk.2015.NPA}).
This is because traditional consideration of the $\alpha$ capture does not take into account the behavior of the wave function inside the internal nuclear region (the corresponding flux is not conserved) which, however, should be defined according to grounds of quantum mechanics.
In terms of our analysis with improved calculations, these quasibound states appear in a complete description of evolution of the compound nucleus by including the contribution from the internal nuclear region.
We describe this evolution in the internal region based on the convergence of flux in the full region.
To completely describe the evolution of the compound-nuclear system, we apply and improve our previous method of the multiple internal reflections \cite{Maydanyuk.2015.NPA} (see also Refs.~\cite{Maydanyuk.MIR.all}).
Advances of our method are
(1) a clear picture of formation of the compound nucleus and its disintegration,
(2) a detailed quantum description of compound-nucleus evolution,
(3) tests of quantum mechanics (not realized in other approaches), and
(4) a high accuracy of calculations (not achieved in other approaches).
Another important issue of our method is that we generalize the idea of Gamow, which has widely been applied for nuclear decay and capture problems (based on tunneling through the barrier, and internal oscillations inside nucleus) to our formalism.
With this, we find a new additional factor.
This factor describes space distribution of the $\alpha$ particle inside nuclear region.
However, we find no discussions of this in previous papers on the topic.

We demonstrate peculiarities of our method through the capture reaction of $\alpha + ^{44}{\rm Ca}$.
In this reaction we predict quasibound energy levels (see Table~\ref{table.3.2.2}), and show that inclusion of evolution of the compound nucleus with possible fusion
allows us to essentially increase the agreement between theory and experimental data.
%
%
This can be seen from Fig.~\ref{fig.3.2}, which shows that the calculated cross section for capture of the $\alpha$ particles by $^{44}{\rm Ca}$
agrees very well with experimental data.
The updated data of fusion probabilities for this reaction are shown in Table~\ref{table.3.2.1}), in comparison with our previous results in Ref.~\cite{Maydanyuk.2015.NPA} (see Tables~2, B.3, and F.9 in that paper).

We compared our formalism with the theory of quasistationary states with complex energies in determination of resonant states in scattering and quasistationary
states in the $\alpha$ capture (see, for example, Refs.~\cite{Baz.1971.book,Vertse.1982.CPC,gamow_functions.all}).
%
%
We found the followings.

\begin{enumerate}
\item
The theory of quasistationary states with complex energies could not provide calculations for the cross section of $\alpha$ capture according to the modern models of $\alpha$ capture (see Refs.~\cite{Denisov.2009.ADNDT,Maydanyuk.2015.NPA,Denisov.2005.PHRVA}).
Our approach provides a unified formalism to calculate the penetrabilities with the best accuracy (in order to estimate the cross sections along the modern formalism of $\alpha$ capture), the probabilities of existence of compound nucleus, and to estimate probabilities of fusion (see discussions in Sec.~\ref{sec.3.2.1}).

\item
The quasistationary states (and corresponding energies) calculated for $\alpha$ capture by the theory of quasistationary states at complex energies differ from quasibound states given by our approach (see Fig.~\ref{fig.3.2.3} and explanation in Sec.~\ref{sec.3.2.3}). We add a comparative analysis between these two approaches for the $\alpha + ^{44}{\rm Ca}$ scattering in Sec.~\ref{sec.3.2.2}.

\end{enumerate}
The difference could be explained by the simplest example in which a free particle moving inside the constant potential; these two approaches describe two principally different processes (for the same real energy, but nonzero $\Gamma$-width).
However, calculations for more complicated realistic potentials are based on such a point.
Time of calculation for the two approaches is similar.
For example, the time of calculations for the modulus of the $S$-matrix (and $\Gamma$-width at complex energies) is around 8~sec (at $N=100$, $r_{\rm max}=70$~fm, 500 intervals in real energy region, realistic barrier), 
and the time of calculations for the probability of existence of the compound nucleus $P_{\rm cn}$ in our method is around 10~sec (with the same parameters).

\section*{Acknowledgements
\label{sec.acknowledgements}}

S.~P.~M. thanks the Institute of Modern Physics of Chinese Academy of Sciences for its warm hospitality and support.
L.-P.~Z. thanks the China Scholarship Council for its support.
This work has been supported by
the Major State Basic Research Development Program in China (No. 2015CB856903),
the National Natural Science Foundation of China (Grant No. 11575254 and 11175215), and
the Chinese Academy of Sciences fellowships for researchers from developing countries (No. 2014FFJA0003).
\end{document}